\documentclass[11pt, a4paper]{article}
\usepackage{jheppub}

\usepackage{tikz}
\definecolor{maroon}{RGB}{128,0,0}
\definecolor{navy}{RGB}{0,0,128}
\definecolor{midnightblue}{RGB}{0,51,102}
\definecolor{darkblue}{RGB}{0,0,139}

\usepackage{relsize}
\usepackage[caption=false,labelformat=simple]{subfig}

\usetikzlibrary{calc, positioning, snakes, arrows.meta}
\tikzset{
  baseline={([yshift=-0.75ex]current bounding box.center)},
  zerosep/.style={inner sep=0pt, outer sep=0pt, minimum size=0pt},
  node distance=8pt,
  align at top/.style={baseline=(current bounding box.north)},
}

\usetikzlibrary{decorations,decorations.markings}
\tikzset{
  strike out/.style={
    postaction=decorate,
    decoration={
      markings,
      mark=at position 0.5 with {
        \draw[-] (-2pt, -3pt) -- (2pt, 3pt);
      }
    }
  }
}

\tikzset{
vev/.style={strike out},
pdvev/.style={snake=snake},
}

\newcommand{\hf}{\mathfrak{h}}

\newcommand{\id}{\mathop{\mathrm{id}}\nolimits}
\newcommand{\sgn}{\mathop{\mathrm{sgn}}\nolimits}

\newcommand{\ket}[1]{|#1\rangle}

\newcommand{\abs}[1]{\lvert #1 \rvert}

\newcommand{\Bigabs}[1]{\Bigl\lvert #1 \Bigr\rvert}

\newcommand{\diag}{\mathop{\mathrm{diag}}\nolimits}

\renewcommand{\Im}{\mathop{\mathrm{Im}}\nolimits}
\renewcommand{\Re}{\mathop{\mathrm{Re}}\nolimits}

\newcommand{\Tr}{\mathop{\mathrm{Tr}}\nolimits}
\newcommand{\End}{\mathop{\mathrm{End}}\nolimits}

\newcommand{\SU}{\mathrm{SU}}

\newcommand{\SL}{\mathrm{SL}}
\newcommand{\slf}{\mathfrak{sl}}

\newcommand{\U}{\mathrm{U}}

\newcommand{\iso}{\cong}

\newcommand{\Z}{\mathbb{Z}}

\newcommand{\R}{\mathbb{R}}
\newcommand{\C}{\mathbb{C}}

\renewcommand{\P}{\mathbb{P}}
\newcommand{\T}{\mathbb{T}}

\let\nc\newcommand
\let\renc\renewcommand
\nc{\wbar}{\overline}
\let\td\tilde
\let\wtd\widetilde
\let\wht\widehat
\let\mcl\mathcal

\nc{\ab}{{\bar{a}}} \nc{\at}{\tilde{a}} \nc{\ah}{\hat{a}}
\nc{\bb}{{\bar{b}}} \nc{\bt}{\tilde{b}} \nc{\bh}{\hat{b}}
\nc{\cb}{{\bar{c}}} \nc{\ct}{\tilde{c}} 
\nc{\db}{{\bar{d}}} \nc{\dt}{\tilde{d}} \renc{\dh}{\hat{d}}
\nc{\eb}{{\bar{e}}} \nc{\et}{\tilde{e}} \nc{\eh}{\hat{e}}
\nc{\fb}{{\bar{f}}} \nc{\ft}{\tilde{f}} \nc{\fh}{\hat{f}}
\nc{\gb}{{\bar{g}}} \nc{\gt}{\tilde{g}} \nc{\gh}{\hat{g}}
\nc{\hb}{{\bar{h}}} \nc{\hh}{\hat{h}} 
\nc{\ib}{{\bar{\imath}}} \nc{\ih}{\hat{\imath}} 
\nc{\jb}{{\bar{\jmath}}} \nc{\jt}{\tilde{\jmath}} \nc{\jh}{\hat{\jmath}}
\nc{\kb}{{\bar{k}}} \nc{\kt}{\tilde{k}} \nc{\kh}{\hat{k}}
\nc{\lb}{{\bar{l}}} \nc{\lt}{\tilde{l}} \nc{\lh}{\hat{l}}
\nc{\mb}{{\bar{m}}} \nc{\mt}{\tilde{m}} \nc{\mh}{\hat{m}}
\nc{\nb}{{\bar{n}}} \nc{\nt}{\tilde{n}} \nc{\nh}{\hat{n}}
\nc{\ob}{{\bar{o}}} \nc{\ot}{\tilde{o}} \nc{\oh}{\hat{o}}
\nc{\pb}{{\bar{p}}} \nc{\pt}{\tilde{p}} \nc{\ph}{\hat{p}}
\nc{\qb}{{\bar{q}}} \nc{\qt}{\tilde{q}} \nc{\qh}{\hat{q}}
\nc{\rb}{{\bar{r}}} \nc{\rt}{\tilde{r}} \nc{\rh}{\hat{r}}
\renc{\sb}{{\bar{s}}} \nc{\st}{\tilde{s}} \nc{\sh}{\hat{s}}
\nc{\tb}{{\bar{t}}} \renc{\th}{\hat{t}} 
\nc{\ub}{{\bar{u}}} \nc{\ut}{\tilde{u}} \nc{\uh}{\hat{u}}
\nc{\vb}{{\bar{v}}} \nc{\vt}{\tilde{v}} \nc{\vh}{\hat{v}}
\nc{\wb}{{\bar{w}}} \nc{\wt}{\tilde{w}} \nc{\wh}{\hat{w}}
\nc{\xb}{{\bar{x}}} \nc{\xt}{\tilde{x}} \nc{\xh}{\hat{x}}
\nc{\yb}{{\bar{y}}} \nc{\yt}{\tilde{y}} \nc{\yh}{\hat{y}}
\nc{\zb}{{\bar{z}}} \nc{\zt}{\tilde{z}} \nc{\zh}{\hat{z}}

\nc{\Ab}{{\wbar{A}}} \nc{\At}{{\wtd{A}}} \nc{\Ah}{{\wht{A}}}
\nc{\Bb}{{\wbar{B}}} \nc{\Bt}{{\wtd{B}}} \nc{\Bh}{{\wht{B}}}
\nc{\Cb}{{\wbar{C}}} \nc{\Ct}{{\wtd{C}}} \nc{\Ch}{{\wht{C}}}
\nc{\Db}{{\wbar{D}}} \nc{\Dt}{{\wtd{D}}} \nc{\Dh}{{\wht{D}}}
\nc{\Eb}{{\wbar{E}}} \nc{\Et}{{\wtd{E}}} \nc{\Eh}{{\wht{E}}}
\nc{\Fb}{{\wbar{F}}} \nc{\Ft}{{\wtd{F}}} \nc{\Fh}{{\wht{F}}}
\nc{\Gb}{{\wbar{G}}} \nc{\Gt}{{\wtd{G}}} \nc{\Gh}{{\wht{G}}}
\nc{\Hb}{{\wbar{H}}} \nc{\Ht}{{\wtd{H}}} \nc{\Hh}{{\wht{H}}}
\nc{\Ib}{{\bar{I}}} \nc{\It}{{\wtd{I}}} \nc{\Ih}{{\wht{I}}}
\nc{\Jb}{{\bar{J}}} \nc{\Jt}{{\wtd{J}}} \nc{\Jh}{{\wht{J}}}
\nc{\Kb}{{\wbar{K}}} \nc{\Kt}{{\wtd{K}}} \nc{\Kh}{{\wht{K}}}
\nc{\Lb}{{\wbar{L}}} \nc{\Lt}{{\wtd{L}}} \nc{\Lh}{{\wht{L}}}
\nc{\Mb}{{\wbar{M}}} \nc{\Mt}{{\wtd{M}}} \nc{\Mh}{{\wht{M}}}
\nc{\Nb}{{\wbar{N}}} \nc{\Nt}{{\wtd{N}}} \nc{\Nh}{{\wht{N}}}
\nc{\Ob}{{\wbar{O}}} \nc{\Ot}{{\wtd{O}}} \nc{\Oh}{{\wht{O}}}
\nc{\Pb}{{\wbar{P}}} \nc{\Pt}{{\wtd{P}}} \nc{\Ph}{{\wht{P}}}
\nc{\Qb}{{\wbar{Q}}} \nc{\Qt}{{\wtd{Q}}} \nc{\Qh}{{\wht{Q}}}
\nc{\Rb}{{\wbar{R}}} \nc{\Rt}{{\wtd{R}}} \nc{\Rh}{{\wht{R}}}
\nc{\Sb}{{\wbar{S}}} \nc{\St}{{\wtd{S}}} \nc{\Sh}{{\wht{S}}}
\nc{\Tb}{{\wbar{T}}} \nc{\Tt}{{\wtd{T}}} \nc{\Th}{{\wht{T}}}
\nc{\Ub}{{\wbar{U}}} \nc{\Ut}{{\wtd{U}}} \nc{\Uh}{{\wht{U}}}
\nc{\Vb}{{\wbar{V}}} \nc{\Vt}{{\wtd{V}}} \nc{\Vh}{{\wht{V}}}
\nc{\Wb}{{\wbar{W}}} \nc{\Wt}{{\wtd{W}}} \nc{\Wh}{{\wht{W}}}
\nc{\Xb}{{\wbar{X}}} \nc{\Xt}{{\wtd{X}}} \nc{\Xh}{{\wht{X}}}
\nc{\Yb}{{\wbar{Y}}} \nc{\Yt}{{\wtd{Y}}} \nc{\Yh}{{\wht{Y}}}
\nc{\Zb}{{\wbar{Z}}} \nc{\Zt}{{\wtd{Z}}} \nc{\Zh}{{\wht{Z}}}

\nc{\CA}{{\mcl{A}}} \nc{\CAb}{{\wbar{\CA}}} \nc{\CAt}{{\wtd{\CA}}} \nc{\CAh}{{\wht{\CA}}}
\nc{\CB}{{\mcl{B}}} \nc{\CBb}{{\wbar{\CB}}} \nc{\CBt}{{\wtd{\CB}}} \nc{\CBh}{{\wht{\CB}}}
\nc{\CC}{{\mcl{C}}} \nc{\CCb}{{\wbar{\CC}}} \nc{\CCt}{{\wtd{\CC}}} \nc{\CCh}{{\wht{\CC}}}
\nc{\cD}{{\mcl{D}}} \nc{\cDb}{{\wbar{\cD}}} \nc{\cDt}{{\wtd{\cC}}} \nc{\cDh}{{\wht{\cD}}}
\nc{\CE}{{\mcl{E}}} \nc{\CEb}{{\wbar{\CE}}} \nc{\CEt}{{\wtd{\CE}}} \nc{\CEh}{{\wht{\CE}}}
\nc{\CF}{{\mcl{F}}} \nc{\CFb}{{\wbar{\CF}}} \nc{\CFt}{{\wtd{\CF}}} \nc{\CFh}{{\wht{\CF}}}
\nc{\CG}{{\mcl{G}}} \nc{\CGb}{{\wbar{\CG}}} \nc{\CGt}{{\wtd{\CG}}} \nc{\CGh}{{\wht{\CG}}}
\nc{\CH}{{\mcl{H}}} \nc{\CHb}{{\wbar{\CH}}} \nc{\CHt}{{\wtd{\CH}}} \nc{\CHh}{{\wht{\CH}}}
\nc{\CI}{{\mcl{I}}} \nc{\CIb}{{\wbar{\CI}}} \nc{\CIt}{{\wtd{\CI}}} \nc{\CIh}{{\wht{\CI}}}
\nc{\CJ}{{\mcl{J}}} \nc{\CJb}{{\wbar{\CJ}}} \nc{\CJt}{{\wtd{\CJ}}} \nc{\CJh}{{\wht{\CJ}}}
\nc{\CK}{{\mcl{K}}} \nc{\CKb}{{\wbar{\CK}}} \nc{\CKt}{{\wtd{\CK}}} \nc{\CKh}{{\wht{\CK}}}
\nc{\CL}{{\mcl{L}}} \nc{\CLb}{{\wbar{\CL}}} \nc{\CLt}{{\wtd{\CL}}} \nc{\CLh}{{\wht{\CL}}}
\nc{\CM}{{\mcl{M}}} \nc{\CMb}{{\wbar{\CM}}} \nc{\CMt}{{\wtd{\CM}}} \nc{\CMh}{{\wht{\CM}}}
\nc{\CN}{{\mcl{N}}} \nc{\CNb}{{\wbar{\CN}}} \nc{\CNt}{{\wtd{\CN}}} \nc{\CNh}{{\wht{\CN}}}
\nc{\CO}{{\mcl{O}}} \nc{\COb}{{\wbar{\CO}}} \nc{\COt}{{\wtd{\CO}}} \nc{\COh}{{\wht{\CO}}}
\nc{\CP}{{\mcl{P}}} \nc{\CPb}{{\wbar{\CP}}} \nc{\CPt}{{\wtd{\CP}}} \nc{\CPh}{{\wht{\CP}}}
\nc{\CQ}{{\mcl{Q}}} \nc{\CQb}{{\wbar{\CQ}}} \nc{\CQt}{{\wtd{\CQ}}} \nc{\CQh}{{\wht{\CQ}}}
\nc{\CR}{{\mcl{R}}} \nc{\CRb}{{\wbar{\CR}}} \nc{\CRt}{{\wtd{\CR}}} \nc{\CRh}{{\wht{\CR}}}
\nc{\CS}{{\mcl{S}}} \nc{\CSb}{{\wbar{\CS}}} \nc{\CSt}{{\wtd{\CS}}} \nc{\CSh}{{\wht{\CS}}}
\nc{\CT}{{\mcl{T}}} \nc{\CTb}{{\wbar{\CT}}} \nc{\CTt}{{\wtd{\CT}}} \nc{\CTh}{{\wht{\CT}}}
\nc{\CU}{{\mcl{U}}} \nc{\CUb}{{\wbar{\CU}}} \nc{\CUt}{{\wtd{\CU}}} \nc{\CUh}{{\wht{\CU}}}
\nc{\CV}{{\mcl{V}}} \nc{\CVb}{{\wbar{\CV}}} \nc{\CVt}{{\wtd{\CV}}} \nc{\CVh}{{\wht{\CV}}}
\nc{\CW}{{\mcl{W}}} \nc{\CWb}{{\wbar{\CW}}} \nc{\CWt}{{\wtd{\CW}}} \nc{\CWh}{{\wht{\CW}}}
\nc{\CX}{{\mcl{X}}} \nc{\CXb}{{\wbar{\CX}}} \nc{\CXt}{{\wtd{\CX}}} \nc{\CXh}{{\wht{\CX}}}
\nc{\CY}{{\mcl{Y}}} \nc{\CYb}{{\wbar{\CY}}} \nc{\CYt}{{\wtd{\CY}}} \nc{\CYh}{{\wht{\CY}}}
\nc{\CZ}{{\mcl{Z}}} \nc{\CZb}{{\wbar{\CZ}}} \nc{\CZt}{{\wtd{\CZ}}} \nc{\CZh}{{\wht{\CZ}}}

\let\eps\epsilon
\let\ups\upsilon
\let\veps\varepsilon
\let\vtht\vartheta

\let\vsgm\varsigma
\let\vphi\varphi
\let\vrho\varrho

\nc{\alphab}{{\bar{\alpha}}} \nc{\alphat}{{\td{\alpha}}} \nc{\alphah}{{\hat{\alpha}}}
\nc{\betab}{{\bar{\beta}}}   \nc{\betat}{{\td{\beta}}}   \nc{\betah}{{\hat{\beta}}} 
\nc{\gammab}{{\bar{\gamma}}} \nc{\gammat}{{\td{\gamma}}} \nc{\gammah}{{\hat{\gamma}}} 
\nc{\deltab}{{\bar{\delta}}} \nc{\deltat}{{\td{\delta}}} \nc{\deltah}{{\hat{\delta}}} 
\nc{\epsilonb}{{\bar{\eps}}} \nc{\epsilont}{{\td{\eps}}} \nc{\epsilonh}{{\hat{\eps}}} 
\nc{\vepsb}{{\bar{\veps}}}   \nc{\vepst}{{\td{\veps}}}   \nc{\vepsh}{{\hat{\veps}}} 
\nc{\zetab}{{\bar{\zeta}}}   \nc{\zetat}{{\td{\zeta}}}   \nc{\zetah}{{\hat{\zeta}}} 
\nc{\etab}{{\bar{\eta}}}     \nc{\etat}{{\td{\eta}}}     \nc{\etah}{{\hat{\eta}}} 
\nc{\thetab}{{\bar{\theta}}} \nc{\thetat}{{\td{\theta}}} \nc{\thetah}{{\hat{\theta}}} 
\nc{\vthetab}{{\bar{\vtht}}} \nc{\vthetat}{{\td{\vtht}}} \nc{\vthetah}{{\hat{\vtht}}} 
\nc{\lambdab}{{\bar{\lambda}}} \nc{\lambdat}{{\td{\lambda}}} \nc{\lambdah}{{\hat{\lambda}}} 
\nc{\iotab}{{\bar{\iota}}}   \nc{\iotat}{{\td{\iota}}}   \nc{\iotah}{{\hat{\iota}}} 
\nc{\kappab}{{\bar{\kappa}}} \nc{\kappat}{{\td{\kappa}}} \nc{\kappah}{{\hat{\kappa}}} 
\nc{\lmdb}{{\bar{\lmd}}}     \nc{\lmdt}{{\td{\lmd}}}     \nc{\lmdh}{{\hat{\lmd}}} 
\nc{\mub}{{\bar{\mu}}}       \nc{\mut}{{\td{\mu}}}       \nc{\muh}{{\hat{\mu}}} 
\nc{\nub}{{\bar{\nu}}}       \nc{\nut}{{\td{\nu}}}       \nc{\nuh}{{\hat{\nu}}} 
\nc{\xib}{{\bar{\xi}}}       \nc{\xit}{{\td{\xi}}}       \nc{\xih}{{\hat{\xi}}} 
\nc{\pib}{{\bar{\pi}}}       \nc{\pit}{{\td{\pi}}}       \nc{\pih}{{\hat{\pi}}} 
\nc{\vpib}{{\bar{\vpi}}}     \nc{\vpit}{{\td{\vpi}}}     \nc{\vpih}{{\hat{\vpi}}} 
\nc{\rhob}{{\bar{\rho}}}     \nc{\rhot}{{\td{\rho}}}     \nc{\rhoh}{{\hat{\rho}}} 
\nc{\vrhob}{{\bar{\vrho}}}   \nc{\vrhot}{{\td{\vrho}}}   \nc{\vrhoh}{{\hat{\vrho}}} 
\nc{\sigmab}{{\bar{\sigma}}} \nc{\sigmat}{{\td{\sigma}}} \nc{\sigmah}{{\hat{\sigma}}} 
\nc{\vsigmab}{{\bar{\vsgm}}} \nc{\vsigmat}{{\td{\vsgm}}} \nc{\vsigmah}{{\hat{\vsgm}}} 
\nc{\taub}{{\bar{\tau}}}     \nc{\taut}{{\td{\tau}}}     \nc{\tauh}{{\hat{\tau}}} 
\nc{\upsb}{{\bar{\ups}}} \nc{\upst}{{\td{\ups}}} \nc{\upsh}{{\hat{\ups}}} 
\nc{\phib}{{\bar{\phi}}}     \nc{\phit}{{\td{\phi}}}     \nc{\phih}{{\hat{\phi}}} 
\nc{\varphib}{{\bar{\vphi}}}   \nc{\varphit}{{\td{\vphi}}}   \nc{\varphih}{{\hat{\vphi}}} 
\nc{\chib}{{\bar{\chi}}}     \nc{\chit}{{\td{\chi}}}     \nc{\chih}{{\hat{\chi}}} 
\nc{\psib}{{\bar{\psi}}}     \nc{\psit}{{\td{\psi}}}     \nc{\psih}{{\hat{\psi}}} 
\nc{\omegab}{{\bar{\omega}}} \nc{\omegat}{{\td{\omega}}} \nc{\omegah}{{\hat{\omega}}} 

\nc{\Gammab}{{\wbar{\Gamma}}}     \nc{\Gammat}{{\wtd{\Gamma}}}     \nc{\Gammah}{{\wht{\Gamma}}}
\nc{\Deltab}{{\wbar{\Delta}}}     \nc{\Deltat}{{\wtd{\Delta}}}     \nc{\Deltah}{{\wht{\Delta}}}
\nc{\Thetab}{{\wbar{\Theta}}}     \nc{\Thetat}{{\wtd{\Theta}}}     \nc{\Thetah}{{\wht{\Theta}}}
\nc{\Lambdab}{{\wbar{\Lambda}}}   \nc{\Lambdat}{{\wtd{\Lambda}}}   \nc{\Lambdah}{{\wht{\Lambda}}}
\nc{\Xib}{{\wbar{\Xi}}}           \nc{\Xit}{{\wtd{\Xi}}}           \nc{\Xih}{{\wht{\Xi}}}
\nc{\Pib}{{\wbar{\Pi}}}           \nc{\Pit}{{\wtd{\Pi}}}           \nc{\Pih}{{\wht{\Pi}}}
\nc{\Sigmab}{{\wbar{\Sigma}}}     \nc{\Sigmat}{{\wtd{\Sigma}}}     \nc{\Sigmah}{{\wht{\Sigma}}}
\nc{\Upsilonb}{{\wbar{\Upsilon}}} \nc{\Upsilont}{{\wtd{\Upsilon}}} \nc{\Upsilonh}{{\wht{\Upsilon}}}
\nc{\Phib}{{\wbar{\Phi}}}         \nc{\Phit}{{\wtd{\Phi}}}         \nc{\Phih}{{\wht{\Phi}}}
\nc{\Psib}{{\wbar{\Psi}}}         \nc{\Psit}{{\wtd{\Psi}}}         \nc{\Psih}{{\wht{\Psi}}}
\nc{\Omegab}{{\wbar{\Omega}}}     \nc{\Omegat}{{\wtd{\Omega}}}     \nc{\Omegah}{{\wht{\Omega}}}

\newcommand{\rmd}{\mathrm{d}}

\newlength{\qsep}
\usetikzlibrary{decorations.markings, arrows, intersections}
\tikzset{
  x=\qsep, y=\qsep,  font=\smaller,
  ->-/.style={decoration={
      markings, mark=at position #1 with
      {\arrow{>}}},postaction={decorate}},
  -<-/.style={decoration={
      markings, mark=at position #1 with
      {\arrow{<}}},postaction={decorate}},
  node/.style={draw, fill=white, shape=circle, minimum size=7pt, inner
    sep=0pt},
  gnode/.style={node},
  dgnode/.style={node, densely dashed},
  ggnode/.style={node, double},
  fnode/.style={node, shape=rectangle},
  tnode/.style={fnode, double, minimum size=12pt},
  q-/.style={-},
  q->/.style={->,  >=stealth, shorten >=1pt, font=\smaller[2]},
  q<-/.style={q->, <-, shorten >=0pt, shorten <=1pt},
  eq-/.style={double, double distance=2pt},
}

\tikzset{
  r-/.style={-, thick},
  r->/.style={r-, ->},
  r<-/.style={r-, <-},
  r->-/.style={->-=#1, thick},
}

\tikzset{
  z-/.style={-, >={Classical TikZ Rightarrow[length=2pt]}},
  z->/.style={z-, ->},
  z<-/.style={z->, <-},
  dr-/.style={-, densely dashed, line width=0.6pt, >={Classical TikZ
      Rightarrow[length=2.5pt]}},
  dr->/.style={dr-, ->},
  dr<-/.style={dr-, <-},
  dtr-/.style={-, double, double distance=1pt, >={Classical TikZ
      Rightarrow[length=3pt]}},
  dtr->/.style={dtr-, ->},
  dtr<-/.style={dtr-, <-},
  tr-/.style={-, line width=1pt, >={Classical TikZ
      Rightarrow[length=3pt]}},
  tr->/.style={tr-, ->},
  tr<-/.style={tr-, <-},
  wz-/.style={z-, densely dotted}, 
  wz->/.style={wz-, ->},
  wz<-/.style={wz-, <-},
  shaded/.style={fill=black!20},
  lshaded/.style={fill=midnightblue!20},
  dshaded/.style={fill=midnightblue!50},
  frame/.style={draw=olive},
  ws/.style={fill=olive!5},
  boundary/.style={draw=maroon},
  fboundary/.style={boundary, double},
}

\tikzset{
  minp/.style={draw, shape=circle, minimum size=3pt, inner 
    sep=0pt, font=\tiny},
  maxp/.style={minp, double, double distance=1pt, fill=white, minimum
    size=5pt},
}
\newcommand{\IV}{\mathcal{I}_{\mathrm{V}}}
\newcommand{\IB}{\mathcal{I}_{\mathrm{B}}}

\newcommand{\V}{\mathbb{V}}

\newcommand{\RB}{R^{\text{B}}}
\newcommand{\LB}{L^{\text{B}}}

\newcommand{\RF}{R^{\text{F}}}
\newcommand{\LF}{L^{\text{F}}}
\newcommand{\LFt}{\Lt^{\text{F}}}
\newcommand{\RBS}{R^{\text{BSDS}}}

\newcommand{\RBt}{\Rt^{\text{B}}}

\newcommand{\RFb}{\Rb^{\text{F}}}
\newcommand{\RBSt}{\Rt^{\text{BS}}}

\newcommand{\dprod}{\displaystyle\prod}

\newcommand{\iu}{\mathrm{i}}

\title{Surface defects and elliptic quantum groups}

\author{Junya Yagi}

\emailAdd{junya.yagi@fuw.edu.pl}

\affiliation{Faculty of Physics, University of Warsaw\\
  ul.\ Pasteura 5, 02--093 Warsaw, Poland}

\abstract{A brane construction of an integrable lattice model is
  proposed.  The model is composed of Belavin's R-matrix, Felder's
  dynamical R-matrix, the Bazhanov--Sergeev--Derkachov--Spiridonov
  R-operator and some intertwining operators.  This construction
  implies that a family of surface defects act on supersymmetric
  indices of four-dimensional $\CN = 1$ supersymmetric field theories
  as transfer matrices related to elliptic quantum groups.}
 

\arxivnumber{}

\preprint{}

\renewcommand{\Delta}{T}

\begin{document}
\maketitle

\section{Introduction}

The aim of this paper is twofold.  It is an attempt to offer a fresh
perspective on integrable lattice models of elliptic type, such as
Baxter's eight-vertex model~\cite{Baxter:1971cr, Baxter:1972hz}, by
embedding them into string theory.  At the same time, it seeks to
provide a new approach to the study of four-dimensional supersymmetric
field theories by connecting their surface operators to those models.

More specifically, I propose a brane construction of an integrable
lattice model that is composed of Belavin's
R-matrix~\cite{Belavin:1981ix}, Felder's dynamical
R-matrix~\cite{Felder:1994pb, Felder:1994be}, the
Bazhanov--Sergeev--Derkachov--Spiridonov (BSDS)
R-operator~\cite{Bazhanov:2010kz, Bazhanov:2011mz, Derkachov:2012iv}
and intertwining operators between the first two
R-matrices~\cite{Baxter:1972wf, MR908997}.  This construction allows
us to map a family of surface defects in $\CN = 1$ supersymmetric
field theories to transfer matrices related to Felder's elliptic
quantum groups.

The proposal builds on a recent development~\cite{Maruyoshi:2016caf}
in the correspondence between $\CN = 1$ supersymmetric field theories
and integrable lattice models~\cite{Spiridonov:2010em,
  Yamazaki:2012cp, Yagi:2015lha, Yagi:2016oum}.  Therefore, let me
first review the relevant results.

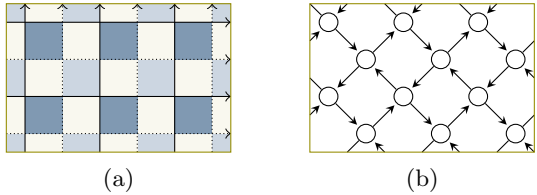
\begin{figure}
  \centering
  \subfloat[\label{fig:BT-D}]{
    \begin{tikzpicture}[scale=1]
      \fill[lshaded] (0,0) rectangle (3, 2);

      \begin{scope}[shift={(0.25,0)}]
        \fill[ws] (0,0) rectangle (0.5,2);
        \fill[ws] (1,0) rectangle (1.5,2);
        \fill[ws] (2,0) rectangle (2.5,2);
      \end{scope}

      \begin{scope}[shift={(0,0.25)}]
        \fill[ws] (0,0) rectangle (3,0.5);
        \fill[ws] (0,1) rectangle (3,1.5);
      \end{scope}

      \begin{scope}[shift={(0.25,0.25)}]
        \fill[dshaded] (0,0) rectangle (0.5,0.5);
        \fill[dshaded] (1,0) rectangle (1.5,0.5);
        \fill[dshaded] (2,0) rectangle (2.5,0.5);
        \fill[dshaded] (0,1) rectangle (0.5,1.5);
        \fill[dshaded] (1,1) rectangle (1.5,1.5);
        \fill[dshaded] (2,1) rectangle (2.5,1.5);
      \end{scope}

      \begin{scope}[shift={(0.25,0)}]
        \draw[z->] (0,0) -- (0,2);
        \draw[wz->] (0.5,0) -- (0.5,2);
        \draw[z->] (1,0) -- (1,2);
        \draw[wz->] (1.5,0) -- (1.5,2);
        \draw[z->] (2,0) -- (2,2);
        \draw[wz->] (2.5,0) -- (2.5,2);
      \end{scope}

      \begin{scope}[shift={(0,0.25)}]
        \draw[wz->] (0,0) -- (3,0);
        \draw[z->] (0,0.5) -- (3,0.5);
        \draw[wz->] (0,1) -- (3,1);
        \draw[z->] (0,1.5) -- (3,1.5);
      \end{scope}

      \draw[frame] (0,0) rectangle (3, 2);
    \end{tikzpicture}
  }
  \qquad
  \subfloat[\label{fig:BT-D-quiver}]{
    \begin{tikzpicture}[scale=1]
      \begin{scope}[shift={(0.25, 0.25)}]
        \node[gnode] (i1) at (0.5, 0) {};
        \node[gnode] (i2) at (1.5, 0) {};
        \node[gnode] (i3) at (2.5, 0) {};
        \node[gnode] (j1) at (0.5, 1) {};
        \node[gnode] (j2) at (1.5, 1) {};
        \node[gnode] (j3) at (2.5, 1) {};

        \node[gnode] (k1) at (0, 0.5) {};
        \node[gnode] (l1) at (1, 0.5) {};
        \node[gnode] (m1) at (2, 0.5) {};
        \node[gnode] (k2) at (0, 1.5) {};
        \node[gnode] (l2) at (1, 1.5) {};
        \node[gnode] (m2) at (2, 1.5) {};
      \end{scope}

      \draw[q->] (i1) -- (l1);
      \draw (i1) -- ++(-135:{0.25*sqrt(2)} );
      \draw[q<-] (i1) -- ++(-45:{0.25*sqrt(2)} );
      \draw[q->] (i2) -- (m1);
      \draw (i2) -- ++(-135:{0.25*sqrt(2)} );
      \draw[q<-] (i2) -- ++(-45:{0.25*sqrt(2)} );
      \draw (i3) -- ++(-135:{0.25*sqrt(2)} );
      \draw[q<-] (i3) -- ++(-45:{0.25*sqrt(2)} );
      \draw (i3) -- ++(45:{0.25*sqrt(2)} );
      \draw[q->] (k1) -- (i1);
      \draw[q<-] (k1) -- ++(-135:{0.25*sqrt(2)} );
      \draw (k1) -- ++(135:{0.25*sqrt(2)} );
      \draw[q->] (l1) -- (i2);
      \draw[q->] (l1) -- (j1);
      \draw[q->] (m1) -- (i3);
      \draw[q->] (m1) -- (j2);
      \draw[q->] (j1) -- (k1);
      \draw[q->] (j1) -- (l2);
      \draw[q->] (j2) -- (m2);
      \draw[q->] (j2) -- (l1);
      \draw[q->] (j3) -- (m1);
      \draw (j3) -- ++(45:{0.25*sqrt(2)} );
      \draw[q<-] (j3) -- ++(-45:{0.25*sqrt(2)} );
      \draw[q->] (k2) -- (j1);
      \draw[q<-] (k2) -- ++(45:{0.25*sqrt(2)} );
      \draw (k2) -- ++(135:{0.25*sqrt(2)} );
      \draw[q<-] (k2) -- ++(-135:{0.25*sqrt(2)} );
      \draw[q->] (l2) -- (j2);
      \draw (m2) -- ++(135:{0.25*sqrt(2)} );
      \draw[q<-] (l2) -- ++(45:{0.25*sqrt(2)} );
      \draw (l2) -- ++(135:{0.25*sqrt(2)} );
      \draw[q->] (m2) -- (j3);
      \draw[q<-] (m2) -- ++(45:{0.25*sqrt(2)} );

      \draw[frame] (0,0) rectangle (3, 2);
    \end{tikzpicture}
  }
  \caption{(a) A periodic lattice colored in a checkerboard-like
    pattern.  (b) The quiver associated with the lattice.  Each node
    is an $\SU(N)$ gauge group and each arrow is a matter multiplet.}
  \label{fig:BT}
\end{figure}

One side of the correspondence is a class of theories realized by
certain configurations of 5-branes in string theory, referred to as
brane tilings~\cite{Hanany:2005ve, Franco:2005rj}.  A theory in this
class has multiple $\SU(N)$ gauge and flavor groups as well as matter
fields transforming in bifundamental representations under these
groups, where $N$ is an integer fixed by the brane configuration.  It
is an example of a quiver gauge theory; its field content can be
encapsulated in a planar quiver diagram.  The quiver itself is
specified by a square lattice whose faces are colored in a
checkerboard-like pattern, which encodes the topology of the 5-branes
interwoven in a ten-dimensional spacetime.  The lattice has two kinds
of ``black'' faces, either light shaded or dark shaded, and two shaded
faces sharing a vertex is required to be of different kinds.
Fig.~\ref{fig:BT} shows an example of such a lattice and the
associated quiver.

The other side of the correspondence is the Bazhanov--Sergeev
model~\cite{Bazhanov:2010kz, Bazhanov:2011mz}, defined on the same
tricolor checkerboard lattice.  To each unshaded face is assigned a
continuous spin variable that takes values in a maximal torus
of~$\SU(N)$.  The Boltzmann weight, or R-operator, of the model is an
integral operator involving elliptic gamma functions.  It solves the
Yang--Baxter equation, ensuring that the model is integrable.

In~\cite{Spiridonov:2010em, Yamazaki:2012cp}, it was discovered that
the supersymmetric index of the gauge theory, formulated on the
three-sphere~$S^3$, coincides with the partition function of the
lattice model.  Under this correspondence, the gauge and flavor groups
are mapped to the spin sites, while the matters are interpreted as
interactions between spins located at different sites.  The
Yang--Baxter equation translates on the gauge theory side to the
invariance of the index under Seiberg duality~\cite{Seiberg:1994pq},
which relates two theories describing the same infrared physics.

As elucidated in~\cite{Yagi:2015lha}, what underlies this
correspondence is the structure of a two-dimensional topological
quantum field theory (TQFT), equipped with line operators that are
localized in an extra dimension.  From the fact that the
supersymmetric index is invariant under continuous deformations of the
checkerboard lattice, one deduces that it is captured by a correlation
function of line operators in a TQFT.  A general argument in
open/closed TQFTs~\cite{Yagi:2016oum} then shows that the correlation
function equals the partition function of a lattice model, the
Bazhanov--Sergeev model in this case.  Finally, the TQFT has a hidden
extra dimension that emerges via string dualities, and its existence
implies the integrability of the model~\cite{Costello:2013zra,
  Costello:2013sla}.

\begin{figure}
  \centering
  \subfloat[\label{fig:D3-line}]{
    \begin{tikzpicture}[scale=1]
      \clip (0,0) rectangle (3,1);

      \fill[lshaded] (0,0) rectangle (3,2);

      \begin{scope}[shift={(0.25,0)}]
        \fill[ws] (0,0) rectangle (0.5,2);
        \fill[ws] (1,0) rectangle (1.5,2);
        \fill[ws] (2,0) rectangle (2.5,2);
      \end{scope}

      \begin{scope}[shift={(0,0.25)}]
        \fill[ws] (0,0) rectangle (3,0.5);
        \fill[ws] (0,1) rectangle (3,1.5);
      \end{scope}

      \begin{scope}[shift={(0.25,0.25)}]
        \fill[dshaded] (0,0) rectangle (0.5,0.5);
        \fill[dshaded] (1,0) rectangle (1.5,0.5);
        \fill[dshaded] (2,0) rectangle (2.5,0.5);
        \fill[dshaded] (0,1) rectangle (0.5,1.5);
        \fill[dshaded] (1,1) rectangle (1.5,1.5);
        \fill[dshaded] (2,1) rectangle (2.5,1.5);
      \end{scope}

      \begin{scope}[shift={(0.25,0)}]
        \draw[z->] (0,0) -- (0,2);
        \draw[wz->] (0.5,0) -- (0.5,2);
        \draw[z->] (1,0) -- (1,2);
        \draw[wz->] (1.5,0) -- (1.5,2);
        \draw[z->] (2,0) -- (2,2);
        \draw[wz->] (2.5,0) -- (2.5,2);
      \end{scope}

      \begin{scope}[shift={(0,0.25)}]
        \draw[wz->] (0,0) -- (3,0);
        \draw[z->] (0,0.5) -- (3,0.5);
        \draw[wz->] (0,1) -- (3,1);
        \draw[z->] (0,1.5) -- (3,1.5);

        \draw[dr->] (0,0.25) -- (3,0.25);
      \end{scope}
    \end{tikzpicture}
  }
  \qquad
  \subfloat[\label{fig:D3-LB}]{
    \begin{tikzpicture}[scale=0.5]
      \fill[ws] (0,0) rectangle (2,2);

      \fill[dshaded] (0,0) rectangle (0.75,2);
      \fill[dshaded] (1.25,0) rectangle (2,2);
      \draw[dr->] (0,1) -- (2,1);
      \draw[wz->] (0.75,0) -- (0.75,2);
      \draw[z->] (1.25,0) -- (1.25,2);
  \end{tikzpicture}
  }
  \qquad
  \subfloat[\label{fig:D3-LF}]{
    \begin{tikzpicture}[scale=0.5]
      \fill[ws] (0,0) rectangle (0.75,2);
      \fill[ws] (1.25,0) rectangle (2,2);
      \fill[dshaded] (0.75,2) rectangle (1.25,0);

      \draw[dr->] (0,1) -- (2,1);
      \draw[z->] (0.75,0)-- (0.75,2);
      \draw[wz->] (1.25,0) -- (1.25,2);
    \end{tikzpicture}
  }
  \caption{(a) A dashed line representing a D3-brane.  It acts on the
    Bazhanov--Sergeev model as a transfer matrix.  (b) The L-operator
    from which the transfer matrix is constructed.  It is also an
    L-operator for the Belavin model.  (c) An L-operator for Felder's
    R-matrix.  It is obtained from the L-operator for the Belavin
    model by an interchange of the left and right halves.}
  \label{fig:D3}
\end{figure}
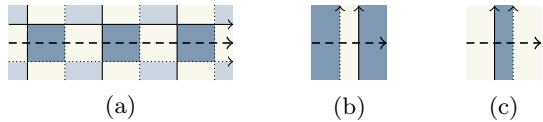

Things get more interesting when we introduce D3-branes that end on
the 5-branes.  With their configurations chosen appropriately, these
additional branes create in the gauge theory surface defects
preserving half of the $\CN = 1$ supersymmetry.  In the checkerboard
lattice they are supported along curves, which we represent by dashed
lines as in Fig.~\ref{fig:D3-line}.  The same reasoning as above,
applied to this situation, leads to the
conclusion~\cite{Maruyoshi:2016caf} that they act on the lattice model
as transfer matrices that consist of so-called L-operators.  An example
of an L-operator is depicted in Fig.~\ref{fig:D3-LB}.

\begin{figure}
  \centering
  \subfloat[\label{fig:Belavin}]{
    \begin{tikzpicture}[scale=1]
      \fill[dshaded] (0,0) rectangle (3, 2);

      \begin{scope}[shift={(0.5,0)}]
        \draw[dr->] (0,0) -- (0,2);
        \draw[dr->] (1,0) -- (1,2);
        \draw[dr->] (2,0) -- (2,2);
      \end{scope}

      \begin{scope}[shift={(0,0.5)}]
        \draw[dr->] (0,0) -- (3,0);
        \draw[dr->] (0,1) -- (3,1);
      \end{scope}

      \draw[frame] (0,0) rectangle (3,2);
    \end{tikzpicture}
  }
  \qquad
  \subfloat[\label{fig:JMO}]{
    \begin{tikzpicture}[scale=1]
      \fill[ws] (0,0) rectangle (3, 2);

      \begin{scope}[shift={(0.5,0)}]
        \draw[dr->] (0,0) -- (0,2);
        \draw[dr->] (1,0) -- (1,2);
        \draw[dr->] (2,0) -- (2,2);
      \end{scope}

      \begin{scope}[shift={(0,0.5)}]
        \draw[dr->] (0,0) -- (3,0);
        \draw[dr->] (0,1) -- (3,1);
      \end{scope}

      \draw[frame] (0,0) rectangle (3,2);
    \end{tikzpicture}
  }
  \caption{(a) A lattice formed by D3-branes in a shaded background
    supports the Belavin model.  (b) The same lattice in an unshaded
    background gives rise to the Jimbo--Miwa--Okado model.}
  \label{fig:Belavin-JMO}
\end{figure}
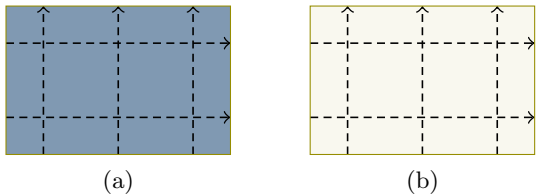

In~\cite{Maruyoshi:2016caf}, the relevant L-operator was identified
for $N = 2$, based on the analysis carried out
in~\cite{Derkachov:2012iv} and independent gauge theory computations.
It was found that this L-operator is essentially Sklyanin's
L-operator~\cite{Sklyanin:1983ig}, and satisfies a defining ``RLL
relation'' not only with the BSDS R-operator but also with Baxter's
R-matrix for the eight-vertex model---a property that uplifts the
well-known relation~\cite{Bazhanov:1989nc} between the chiral Potts
model and the six-vertex model to the elliptic level.  This fact
strongly suggests that for $N = 2$, the eight-vertex model arises when
D3-branes form a lattice in a shaded background, as illustrated in
Fig.~\ref{fig:Belavin}.

In this paper, the above result is generalized in a few interrelated
directions.

First of all, the model realized on the lattice of D3-branes is
identified in the general case $N \geq 2$.  I propose that it is
Belavin's $\Z_N$-symmetric model~\cite{Belavin:1981ix}, which reduces
to the eight-vertex model when $N = 2$.  This proposal is backed up by
the observation that there exists a simultaneous L-operator for the
Belavin model and the Bazhanov--Sergeev model, and it
factorizes~\cite{Sergeev:1992ap, Quano:1992wc, MR1190749} into a pair
of intertwining operators.  The factorization structure is manifest in
the graphical representation of the L-operator, in which a single
intertwining operator corresponds to either the left or right half.

A question then arises as to what the same lattice gives rise to when
it is placed in an unshaded background, as in Fig.~\ref{fig:JMO}.  The
answer should be the model defined by Felder's dynamical R-matrix for
the elliptic quantum group associated with $\slf_N$,%
\footnote{The same conclusion was reached by Kevin Costello through
  consideration of a different physical setup.}
which is the unrestricted $A^{(1)}_{N-1}$ model of Jimbo, Miwa and
Okado~\cite{MR908997, Jimbo:1987mu, Jimbo:1987ra}.  One of the reasons
is that an interchange of the pair of intertwining operators turns the
L-operator for the Belavin model into that for Felder's R-matrix.  The
latter L-operator is graphically represented as in
Fig.~\ref{fig:D3-LF}, and defines a representation of the elliptic
quantum group.

The brane construction therefore unifies three integrable lattice
models, namely the Belavin, Jimbo--Miwa--Okado and Bazhanov--Sergeev
models, with the help of intertwining operators.  The vertex--face
correspondence~\cite{Baxter:1972wf, MR908997} relating Belavin's and
Felder's R-matrices can be understood as Yang--Baxter equations in
this unified model, and as such, admits an interpretation in terms of
brane movements.  The intertwining operators are determined from these
and other Yang--Baxter equations, and shown to lead to transfer
matrices that agree with known results~\cite{Maruyoshi:2016caf,
  Ito:2016fpl} about surface defects in class-$\CS_k$
theories~\cite{Gaiotto:2015usa}.

Lastly, the correspondence with transfer matrices is extended to a
family of surface defects labeled by the irreducible
finite-dimensional representations of $\SU(N)$.  Such a family is
constructed with a little more elaborate brane configurations than the
one described above, which can only handle the vector representation.
This construction is identified with a method to generate arbitrary
irreducible representations on the lattice model side, called the
fusion procedure~\cite{Kulish:1981gi, MR839706}.  The L-operators in
the exterior powers of the vector representation are especially
interesting since their traces produce~\cite{MR1463830} mutually
commuting difference operators known as the Ruijsenaars
operators~\cite{Ruijsenaars:1986pp}.  These traces indeed match the
difference operators that represent the corresponding surface defects
acting on the indices of $\CN = 2$ supersymmetric field
theories~\cite{Bullimore:2014nla}.

The paper is organized as follows.  In section~\ref{sec:R}, we review
the Belavin model, Felder's dynamical R-matrix and the
Jimbo--Miwa--Okado model, and the Bazhanov--Sergeev model.  In
section~\ref{sec:unification}, we construct an integrable lattice
model that unifies the three models introduced in section~\ref{sec:R},
and explain how L-operators constructed from the intertwining
operators provide representations of elliptic quantum groups.  In
section~\ref{sec:SD}, we discuss the brane construction and the
correspondence between surface defects and transfer matrices, and
check the proposal of the paper against gauge theory results.

\section{Integrable lattice models of elliptic type}
\label{sec:R}

In this section we review the three integrable lattice models of
elliptic type relevant to us: the Belavin model, the
Jimbo--Miwa--Okado model defined by Felder's dynamical R-matrix, and
the Bazhanov--Sergeev model.  First we set up notation.

Fix an integer $N \geq 2$ and let $\{e_1, \dotsc, e_N\}$ be the
standard basis of $\C^N$.  The Cartan subalgebra of $\slf_N$ is
$\hf = \{\diag(\lambda_1, \dotsc, \lambda_N) \mid \sum_{i=1}^N
\lambda_i = 0\}$, the space of diagonal traceless complex $N \times N$
matrices.  We make the identification
$ \hf^* \iso \hf \iso \{\lambda \in \C^N \mid \sum_{i=1}^N \lambda_i =
0\} $, where the first isomorphism is induced by the nondegenerate
bilinear form $(\lambda,\mu) = \Tr(\lambda\mu)$ on $\hf$.  The vector
representation $V = \C^N$ of $\slf_N$ has the weight decomposition
$V = \bigoplus_{i=1}^N V[\omega_i]$, with
$\omega_i = e_i - \sum_{j=1}^N e_j/N$ and $V[\omega_i] = \C e_i$.

Setting $z_i = e^{2\pi\iu\lambda_i}$, we obtain an $N$-tuple
$(z_1, \dotsc, z_N)$ of complex variables obeying the constraint
$\prod_{i=1}^N z_i = 1$.  These variables parametrize a maximal torus
of $\SL(N)$.  We denote by $\V$ the space of meromorphic functions on
the maximal torus that are invariant under the action of the Weyl
group, i.e., the space of symmetric meromorphic functions of
$(z_1, \dotsc, z_N)$.  Elements of the tensor product $\V^{\otimes n}$
may be thought of as meromorphic functions of $n$ sets of $N$
variables, symmetric with respect to each set.  We extend
$\V^{\otimes n}$ to mean the space of all such functions.

Furthermore, fix complex parameters $\tau$, $\gamma$ with $\Im\tau$,
$\Im \gamma > 0$, and set $p = e^{2\pi\iu\tau}$ and
$q = e^{2\pi\iu\gamma}$.  The theta function with characteristics is
defined by
\begin{equation}
  \theta\biggl[
  \begin{array}{c}
    a \\ b
  \end{array}
  \biggr]
  (u|\tau)
  =
  \sum_{n=-\infty}^\infty
  e^{\pi\iu (n+a)^2 \tau + 2\pi\iu(n+a)(u+b)}
  \,.
\end{equation}
We will need Jacobi's first theta function and the elliptic gamma
function:
\begin{equation}
  \theta_1(u|\tau)
  =
  -
  \theta\biggl[
  \begin{array}{c}
    1/2 \\ 1/2
  \end{array}
  \biggr]
  (u|\tau)
  \,,
  \qquad
  \Gamma(z;p,q)
  =
  \prod_{m,n=0}^\infty
  \frac{1 - p^{m+1} q^{n+1}/z}{1 - p^m q^n z}
  \,.
\end{equation}
They satisfy $\theta_1(u|\tau) = -\theta_1(-u|\tau)$ and
$\Gamma(z;p,q) = 1/\Gamma(pq/z;p,q)$.  We also introduce
\begin{equation}
  \theta^{(j)}(u|\tau,N)
  =
  \theta\biggl[
  \begin{array}{c}
    1/2 - j/N \\ 1/2
  \end{array}
  \biggr]
  (u|N\tau)
  \,,
  \qquad
  \theta(z;p)
  =
  (z;p)_\infty (p/z;p)_\infty
  \,,
\end{equation}
where we have used the $q$-Pochhammer symbol
$(z;q)_n = \prod_{k=0}^{n-1} (1 - q^k z)$.  The modified theta function
$\theta(z;p)$ is related to $\theta_1(u|\tau)$ by
$\theta_1(u|\tau) = \iu p^{1/8} (p;p)_\infty e^{-\pi\iu u}
\theta(e^{2\pi\iu u};p)$ and satisfies
$\theta(z;p) = \theta(p/z;p) = -z\theta(1/z;p)$.  Moreover, we have
$\Gamma(qz;p,q) = \theta(z;p) \Gamma(z;p,q)$.  In what follows we will
simply write $\theta_1(u)$, $\Gamma(z)$, $\theta^{(j)}(u)$ and
$\theta(z)$ for these functions.

\subsection{Belavin model}

The Belavin model~\cite{Belavin:1981ix} is an integrable lattice model
that generalizes Baxter's eight-vertex model~\cite{Baxter:1971cr,
  Baxter:1972hz}.  It is a vertex model, meaning that its spin
variables reside on the edges and interact at the vertices of the
lattice.  These spins take values in $V$, and the local Boltzmann
weight for a configuration of four spins placed around a vertex is
determined by an operator $\RB \colon \C \to \End(V \otimes V)$, the
R-matrix of the model.

Belavin's R-matrix $\RB(u)$ is a unique $\End(V \otimes V)$-valued
meromorphic function that has simple poles at
$u = -\gamma + \Z + \tau\Z$, satisfies the initial condition
\begin{equation}
  \RB(0) = P\colon v \otimes w \mapsto w \otimes v\,,
\end{equation}
and possesses the $\Z_N$-symmetry
\begin{equation}
  \RB(u)
  = (g \otimes g) \RB(u) (g \otimes g)^{-1}
  = (h \otimes h) \RB(u) (h \otimes h)^{-1}
\end{equation}
and the quasi-periodicity
\begin{align}
  \RB(u+1)
  &=
  (g \otimes \id_{V})^{-1} \RB(u) (g \otimes \id_{V})
  \,,
  \\
  \RB(u+\tau)
  &=
  q^{1-1/N} (h \otimes \id_{V})\RB(u) (h \otimes \id_{V})^{-1}
  \,.
\end{align}
Here $g$, $h \in \End(V)$ are matrices such that
$ge_k = e^{2\pi\iu k/N} e_k$ and $he_k = e_{k+1}$, with
$e_{N+1} = e_1$.  The variable $u$ is called the spectral parameter.

In terms of the matrix elements $\RB(u)_{ij}^{kl}$ defined by
$\RB(u) (e_i \otimes e_j) = \sum_{k,l} \RB(u)_{ij}^{kl} e_k \otimes
e_l$, the R-matrix is given by~\cite{Richey:1986vt}
\begin{equation}
  \RB(u)_{ij}^{kl}  
  =
  \delta_{i+j, k+l}
  \dfrac{\theta_1(\gamma)}{\theta_1(u + \gamma)}
  \dfrac{\theta^{(k-l)}(u + \gamma)}{\theta^{(k - i)}(\gamma)
    \theta^{(i-l)}(u)} 
  \dfrac{\prod_{m=0}^{N-1} \theta^{(m)}(u)}{\prod_{n=1}^{N-1} \theta^{(n)}(0)}
  \,,
\end{equation}
where the indices $i$, $j$, $k$, $l$ are treated modulo $N$.  It can
be shown that $\RB(u)$ satisfies the unitarity relation
\begin{equation}
  \label{eq:RB-unitarity}
  \RB(u) \RB_{21}(-u)  = \id_{V \otimes V}
\end{equation}
and solves the Yang--Baxter equation
\begin{equation}
  \label{eq:RB-YBE}
  \RB_{12}(u_1 - u_2) 
  \RB_{13}(u_1 - u_3) 
  \RB_{23}(u_2 - u_3) 
  =
  \RB_{23}(u_2 - u_3) 
  \RB_{13}(u_1 - u_3) 
  \RB_{12}(u_1 - u_2)
  \,.
\end{equation}
In these equations the subscripts on the R-matrices indicate the space
on which they act, e.g., $\RB_{21}(u) = P \RB(u) P$ and
$\RB_{23}(u) = \id_{V} \otimes \RB(u)$.

We represent the R-matrix graphically by the crossing of two dashed
lines in a dark shaded background:
\begin{equation}
  \label{eq:RB-g}
  \RB(u_1 - u_2)
  =
  \begin{tikzpicture}[scale=0.6, baseline=(x.base)]
    \node (x) at (1,1) {\vphantom{x}};

    \fill[dshaded] (0,0) rectangle (2,2);
    \draw[dr->] (0,1) node[left] {$u_1$} --  (2,1);
    \draw[dr->] (1,0) node[below] {$u_2$} -- (1,2);
  \end{tikzpicture}
  \ .
\end{equation}
Each dashed line carries a spectral parameter, and to each segment of
a dashed line is assigned a copy of $V$.  The R-matrix depends on the
difference of the spectral parameters~$u_1$, $u_2$ of the two lines.
We may think of the right-hand side as depicting the trajectories of
two particles with rapidities $u_1$ and $u_2$, scattering in $1+1$
spacetime dimensions.  The state space of each particle is $V$.  The
matrix element $\RB(u_1 - u_2)_{ij}^{kl}$ is the S-matrix element for
the scattering process with initial state
$\ket{u_1, i} \otimes \ket{u_2, j}$ and final state
$\ket{u_1, k} \otimes \ket{u_2, l}$.

In this graphical notation, the unitarity relation
\eqref{eq:RB-unitarity} is expressed as
\begin{equation}
  \def\pathA{(0,0) to[out=0, in=180] (1,0.5) to[out=0, in=180] (2,0)}
  \def\pathB{(2,0.5) to[in=0,out=180] (1,0) to[in=0,out=180] (0,0.5)}
  \begin{tikzpicture}[baseline=(x.base)]
    \node (x) at (1,0.25) {\vphantom{x}};

    \fill[dshaded] (0,-0.3) rectangle (2,0.8);

    \draw[dr->] node[left] {$u_1$} \pathA;
    \draw[dr<-] \pathB node[left] {$u_2$};
  \end{tikzpicture}
  \ =
  \begin{tikzpicture}[baseline=(x.base)]
    \node (x) at (1,0.25) {\vphantom{x}};

    \fill[dshaded] (0,-0.3) rectangle (2,0.8);

    \draw[dr->] (0,0) node[left] {$u_1$} -- (2,0);
    \draw[dr->] (0,0.5) node[left] {$u_2$} -- (2,0.5);
  \end{tikzpicture}
  \ ,
\end{equation}
whereas the Yang--Baxter equation \eqref{eq:RB-YBE} takes the form of
an equality between two configurations of three dashed lines:
\begin{equation}
  \label{eq:YBE-g}
  \begin{tikzpicture}[scale=0.6, baseline=(x.base)]
    \node (x) at (30:2) {\vphantom{x}};
    
    \fill[dshaded] (0,-0.5) rectangle ({3*sqrt(3)/2},2.5);

    \draw[dr->] (0,0) node[left] {$u_2$} -- ++(30:3);
    \draw[dr->] (0,2) node[left] {$u_1$} -- ++(-30:3);
    \draw[dr->] (-30:1) node[below] {$u_3$} -- ++(0,3);
  \end{tikzpicture}
  \ =
  \begin{tikzpicture}[scale=0.6, baseline=(x.base)]
    \node (x) at (30:1) {\vphantom{x}};
    
    \fill[dshaded] (0,-1) rectangle ({3*sqrt(3)/2},2);

    \draw[dr->] (0,0) node[left] {$u_2$} -- ++(30:3);
    \draw[dr->] (0,1) node[left] {$u_1$} -- ++(-30:3);
    \draw[dr->] (-30:2) node[below] {$u_3$} -- ++(0,3);
  \end{tikzpicture}
  \ .
\end{equation}
The latter relation states that a three-particle scattering process
factorizes into two-particle scattering processes, and the order of
this factorization is immaterial.

Now, consider a lattice constructed from dashed lines in a dark shaded
background, such as the one shown in Fig.~\ref{fig:Belavin}.  The
partition function of the Belavin model on this lattice is defined as
follows.  For each configuration of states on the edges, we define the
corresponding Boltzmann weight to be the product of the R-matrix
elements that arise from the crossings.  Then, the partition function
is the sum of the Boltzmann weights for all possible configurations of
states.  If the lattice is placed on a surface with boundary, then the
states on the edges that intersect the boundary are fixed and not
summed over.

Note that if we reverse the directions of the lines in the
Yang--Baxter equation~\eqref{eq:YBE-g}, we end up with the same set of
pictures.  This means that the transpose $\RBt(u) = \RB(u)^T$ of
$\RB(u)$ is also a solution of the Yang--Baxter equation.  We
represent it by the crossing of two dashed lines in a light shaded
background:
\begin{equation}
  \label{eq:RBt-g}
  \RBt(u_1 - u_2)
  =
  \begin{tikzpicture}[scale=0.6, baseline=(x.base)]
    \node (x) at (1,1) {\vphantom{x}};

    \fill[lshaded] (0,0) rectangle (2,2);
    \draw[dr->] (0,1) node[left] {$u_1$} --  (2,1);
    \draw[dr->] (1,0) node[below] {$u_2$} -- (1,2);
  \end{tikzpicture}
  \ =
  \begin{tikzpicture}[scale=0.6, baseline=(x.base)]
    \node (x) at (1,1) {\vphantom{x}};

    \fill[dshaded] (0,0) rectangle (2,2);
    \draw[dr<-] (0,1) node[left] {$u_1$} --  (2,1);
    \draw[dr<-] (1,0) node[below] {$u_2$} -- (1,2);
  \end{tikzpicture}
  \ .
\end{equation}

When $N = 2$, Belavin's R-matrix reduces to Baxter's R-matrix for the
eight-vertex model.  In this case we have $\RBt(u) = \RB(u)$, hence
there is no distinction between the two kinds of shading.

\subsection{Felder's dynamical R-matrix}
\label{sec:Felder}

Next, we turn to Felder's dynamical R-matrix for the elliptic quantum
group~$E_{\tau, \gamma/2}(\slf_N)$.  This R-matrix first appeared
in~\cite{MR908997} as the Boltzmann weight of a generalization of the
eight-vertex solid-on-solid model~\cite{Baxter:1972wf}.  Later,
Felder~\cite{Felder:1994pb, Felder:1994be} reformulated it as
described here.

Compared to Belavin's, Felder's R-matrix
$\RF(u,\lambda) \in \End(V \otimes V)$ depends on an additional
parameter $\lambda \in \hf^*$, called the dynamical variable.  The
R-matrix satisfies the unitarity relation
\begin{equation}
  \RF(u, \lambda)
  \RF_{21}(-u, \lambda) 
  = \id_{V \otimes V}
\end{equation}
and the dynamical Yang--Baxter equation
\begin{multline}
  \label{eq:dYBE}
  \RF_{12}(u_1 - u_2, \lambda - \gamma h_3) 
  \RF_{13}(u_1 - u_3, \lambda) 
  \RF_{23}(u_2 - u_3, \lambda - \gamma h_1) 
  \\
  =
  \RF_{23}(u_2 - u_3, \lambda) 
  \RF_{13}(u_1 - u_3, \lambda - \gamma h_2) 
  \RF_{12}(u_1 - u_2, \lambda)
  \,.
\end{multline}
Here $h$ stands for the weight of the relevant state; for example,
$\RF_{12}(u_1 - u_2, \lambda - \gamma h_3)$ acts on
$v_1 \otimes v_2 \otimes v_3$ as
$\RF(u_1 - u_2, \lambda - \gamma \mu_3) \otimes \id_{V}$ if the weight
of $v_3$ is $\mu_3$.  The nonzero matrix elements of $\RF(u, \lambda)$
are given by \cite{MR1606760}
\begin{equation}
  \RF(u, \lambda)_{ii}^{ii}
  =
  1
  \,,
  \quad
  \RF(u, \lambda)_{ij}^{ij}
  =
  \frac{\theta_1(u) \theta_1(\lambda_{ij} + \gamma)}
        {\theta_1(u + \gamma) \theta_1(\lambda_{ij})}
  \,,
  \quad
  \RF(u, \lambda)_{ij}^{ji}
  =
  \frac{\theta_1(\gamma) \theta_1(u + \lambda_{ij})}
       {\theta_1(u + \gamma) \theta_1(\lambda_{ij})}
  \,,
\end{equation}
where $i \neq j$ and $\lambda_{ij} = \lambda_i - \lambda_j$.

Graphically, we represent Felder's R-matrix by the crossing of two
dashed lines in an unshaded background:
\begin{equation}
  \label{eq:RF-g}
  \RF(u_1 - u_2, \lambda)
  =
  \begin{tikzpicture}[scale=0.6, baseline=(x.base)]
    \node (x) at (1,1) {\vphantom{x}};

    \fill[ws] (0,0) rectangle (2,2);
    \draw[dr->] (0,1) node[left] {$u_1$} --  (2,1);
    \draw[dr->] (1,0) node[below] {$u_2$} -- (1,2);

    \node at (0.5,1.5) {$\lambda$};
  \end{tikzpicture}
  \ .
\end{equation}
We have marked the upper-left region with $\lambda$ to indicate the
dependence on the dynamical variable.  As before, the unitarity
relation is the equality
\begin{equation}
  \def\pathA{(0,0) to[out=0, in=180] (1,0.5) to[out=0, in=180] (2,0)}
  \def\pathB{(2,0.5) to[in=0,out=180] (1,0) to[in=0,out=180] (0,0.5)}
  \begin{tikzpicture}[baseline=(x.base)]
    \node (x) at (1,0.25) {\vphantom{x}};

    \fill[ws] (0,-0.3) rectangle (2,0.9);

    \draw[dr->] node[left] {$u_1$} \pathA;
    \draw[dr<-] \pathB node[left] {$u_2$};
    
    \node at (1,0.7) {$\lambda$};
  \end{tikzpicture}
  \ =
  \begin{tikzpicture}[baseline=(x.base)]
    \node (x) at (1,0.25) {\vphantom{x}};

    \fill[ws] (0,-0.3) rectangle (2,0.9);

    \draw[dr->] (0,0) node[left] {$u_1$} -- (2,0);
    \draw[dr->] (0,0.5) node[left] {$u_2$} -- (2,0.5);
    
    \node at (1,0.7) {$\lambda$};
  \end{tikzpicture}
  \ .
\end{equation}

To express the dynamical Yang--Baxter equation \eqref{eq:dYBE}
graphically, we demand that across a dashed line segment, the value of
the dynamical variable changes by $\gamma$ times the weight of the
state supported on that segment:
\begin{equation}
  \label{eq:lambda-change}
  \begin{tikzpicture}[scale=0.6, baseline=(x.base)]
    \node (x) at (1,1) {\vphantom{x}};

    \fill[ws] (0,0.3) rectangle (3,2.2);

    \draw[dr->] (0,1.25) -- (3,1.25);

    \node at (1.5,1.75) {$\lambda$};
    \node at (1.5,0.75) {$\lambda - \gamma h$};
  \end{tikzpicture}
  \ .
\end{equation}
Then we have
\begin{equation}
  \label{eq:dYBE-g}
  \begin{tikzpicture}[scale=0.6, baseline=(x.base)]
    \node (x) at (30:2) {\vphantom{x}};

    \fill[ws] (0,-0.5) rectangle ({3*sqrt(3)/2},2.5);

    \draw[dr->] (0,0) node[left] {$u_2$} -- ++(30:3);
    \draw[dr->] (0,2) node[left] {$u_1$} -- ++(-30:3);
    \draw[dr->] (-30:1) node[below] {$u_3$} -- ++(0,3);

    \node (O) at ({sqrt(3)*4/6},1) {};
    \node at ($(O) + (120:{sqrt(3)*9/12})$) {$\lambda$};
  \end{tikzpicture}
  \ =
  \begin{tikzpicture}[scale=0.6, baseline=(x.base)]
    \node (x) at (30:1) {\vphantom{x}};

    \fill[ws] (0,-1) rectangle ({3*sqrt(3)/2},2);

    \draw[dr->] (0,0) node[left] {$u_2$} -- ++(30:3);
    \draw[dr->] (0,1) node[left] {$u_1$} -- ++(-30:3);
    \draw[dr->] (-30:2) node[below] {$u_3$} -- ++(0,3);

    \node (O) at ({sqrt(3)*5/6},0.5) {};
    \node at ($(O) + (120:{sqrt(3)*7/12})$) {$\lambda$};
  \end{tikzpicture}
  \ ,
\end{equation}
just as in the nondynamical case.

Felder's R-matrix has the property
$\RF(u,\lambda) = \RF(u,\lambda + \gamma(h_1 + h_2))$, or
\begin{equation}
  \label{eq:lambda-shift}
  \begin{tikzpicture}[scale=0.6, baseline=(x.base)]
    \node (x) at (1,1) {\vphantom{x}};

    \fill[ws] (0,0) rectangle (2,2);

    \draw[dr->] (0,1) node[left] {$u_1$} -- (2,1);
    \draw[dr->] (1,0) node[below] {$u_2$} -- (1,2);
    
    \node at (0.5,1.5) {$\lambda$};
  \end{tikzpicture}
  \ =
  \begin{tikzpicture}[scale=0.6, baseline=(x.base)]
    \node (x) at (1,1) {\vphantom{x}};

    \fill[ws] (0,0) rectangle (2,2);

    \draw[dr->] (0,1) node[left] {$u_1$} -- (2,1);
    \draw[dr->] (1,0) node[below] {$u_2$} -- (1,2);
    
    \node at (1.5,0.5) {$\lambda$};
  \end{tikzpicture}
  \ .
\end{equation}
Also, the dynamical Yang--Baxter equation holds after the directions
of the lines are reversed:
\begin{equation}
  \begin{tikzpicture}[scale=0.6, baseline=(x.base)]
    \node (x) at (30:2) {\vphantom{x}};

    \fill[ws] (0,-0.5) rectangle ({3*sqrt(3)/2},2.5);

    \draw[dr<-] (0,0) node[left] {$u_2$} -- ++(30:3);
    \draw[dr<-] (0,2) node[left] {$u_1$} -- ++(-30:3);
    \draw[dr<-] (-30:1) node[below] {$u_3$} -- ++(0,3);

    \node (O) at ({sqrt(3)*4/6},1) {};
    \node at ($(O) + (120:{sqrt(3)*9/12})$) {$\lambda$};
  \end{tikzpicture}
  \ =
  \begin{tikzpicture}[scale=0.6, baseline=(x.base)]
    \node (x) at (30:1) {\vphantom{x}};

    \fill[ws] (0,-1) rectangle ({3*sqrt(3)/2},2);

    \draw[dr<-] (0,0) node[left] {$u_2$} -- ++(30:3);
    \draw[dr<-] (0,1) node[left] {$u_1$} -- ++(-30:3);
    \draw[dr<-] (-30:2) node[below] {$u_3$} -- ++(0,3);

    \node (O) at ({sqrt(3)*5/6},0.5) {};
    \node at ($(O) + (120:{sqrt(3)*7/12})$) {$\lambda$};
  \end{tikzpicture}
  \ .
\end{equation}
From these relations it follows that the R-matrix
$\vphantom{R}^{\Theta}\!\RF(u,\lambda) = \RF(u,-\lambda)^T$ also
satisfies the dynamical Yang--Baxter equation.  Note the replacement
$\lambda \to -\lambda$ in the argument.  This is necessary to correct
the way in which the dynamical variable changes since the lines have
been flipped.

In fact, this R-matrix is related to $\RF(u,\lambda)$ by conjugation:
\begin{equation}
  \label{eq:RFt}
  \RF(u,-\lambda)^T
  =
  \Theta_1(\lambda - \gamma h_2) \Theta_2(\lambda)
  \RF(u,\lambda)
  \Theta_1(\lambda)^{-1} \Theta_2(\lambda - \gamma h_1)^{-1}
  \,,
\end{equation}
where the matrix $\Theta(\lambda) \in \End(V)$ is given by
\begin{equation}
  \label{eq:Theta}
  \Theta(\lambda)^i_j
  =
  \delta^i_j \prod_{k (\neq j)} \frac{1}{\theta_1(\lambda_{kj})}
  \,.
\end{equation}
In general, if an R-matrix $R(u,\lambda)$ solves the dynamical
Yang--Baxter equation, then so does
$\vphantom{R}^M \! R(u,\lambda) = M_1(\lambda - \gamma h_2) M_2(\lambda)
R(u,\lambda) M_1(\lambda)^{-1} M_2(\lambda - \gamma h_1)^{-1}$ for any
invertible matrix-valued function $M(\lambda)$.

\subsection{Vertex--face correspondence and Jimbo--Miwa--Okado model}
\label{sec:VFC}

Felder's R-matrix can be obtained from Belavin's by conjugation.  Let
us define a matrix $\Phi(u,\lambda) \in \End(V)$ by
\begin{equation}
  \Phi(u, \lambda)^j_i
  =
  \theta^{(j)}\Bigl(u - N\lambda_i + \frac{N-1}{2}\Bigr)
  \,.
\end{equation}
Then the two R-matrices are related as follows~\cite{Baxter:1972wf,
  MR908997}:
\begin{equation}
  \label{eq:RBPhiPhi}
  \RB(u_1 - u_2)
  \Phi_1(u_1, \lambda)
  \Phi_2(u_2, \lambda + \gamma h_1)
  =
  \Phi_2(u_2, \lambda)
  \Phi_1(u_1, \lambda + \gamma h_2)
  \RF(u_1 - u_2, \lambda)^T
  \,.
\end{equation}
We can rewrite this relation as
\begin{equation}
  \label{eq:RBPsiPsi}
  \RB(u_1 - u_2)
  \Psi_1(u_1,\lambda)
  \Psi_2(u_2, \lambda - \gamma h_1)
  =
  \Psi_2(u_2,\lambda)
  \Psi_1(u_1, \lambda - \gamma h_2)
  \RF(u_1 - u_2, \lambda)
  \,,
\end{equation}
where we have introduced
\begin{equation}
  \Psi(u,\lambda)
  =
  \Phi(u, -\lambda) \Theta(\lambda)
\end{equation}
and used \eqref{eq:RFt} to remove the transpose on $\RF$.

The above relation is called the vertex--face correspondence (or
vertex--IRF transformation) since it transforms a vertex model to an
interaction-round-a-face (IRF) model and vice versa.  An IRF model has
spins placed on the faces, and assigns a local Boltzmann weight to a
configuration of four spins surrounding a vertex.  From the point of
view of the dual lattice, interaction takes place among spins located
round a face.

The model defined by Felder's R-matrix is naturally a vertex model,
but it can also be formulated as an IRF model.  Once the states are
specified on all edges, the values of the dynamical variables living
on the faces are determined by rule~\eqref{eq:lambda-change} up to
overall shifts.  Conversely, a consistent assignment of dynamical
variables to the faces determines the states on the edges completely.
Hence, instead of these states we may think of the dynamical variables
as the spin variables of the model.  The IRF model thus obtained is
the unrestricted $A^{(1)}_{N-1}$ model of Jimbo, Miwa and
Okado~\cite{MR908997, Jimbo:1987mu, Jimbo:1987ra}.

\subsection{Bazhanov--Sergeev model}
\label{sec:BS}

The last integrable lattice model that we review is the
Bazhanov--Sergeev model.  This model was introduced
in~\cite{Bazhanov:2010kz} for $N = 2$ and subsequently extended to
$N \geq 2$ in~\cite{Bazhanov:2011mz}.  It has continuous spin
variables taking values in a maximal torus of $\SU(N)$.  Accordingly,
its Boltzmann weight is given by an infinite-dimensional R-matrix,
which is really an integral operator.

For the purpose of describing the Bazhanov--Sergeev model, it is more
convenient to switch to multiplicative notation, i.e., from $\tau$,
$\gamma$, $\lambda_i$ to $p$, $q$, $z_i$ as defined at the beginning
of this section.  For example, we will write the dynamical
Yang--Baxter equation~\eqref{eq:dYBE} as
\begin{multline}
  \label{eq:dYBE-m}
  \RF_{12}\Bigl(\frac{c_1}{c_2}; q^{-h_3} z\Bigr) 
  \RF_{13}\Bigl(\frac{c_1}{c_3}; z\Bigr) 
  \RF_{23}\Bigl(\frac{c_2}{c_3}; q^{-h_1} z\Bigr) 
  \\
  =
  \RF_{23}\Bigl(\frac{c_2}{c_3}; z\Bigr) 
  \RF_{13}\Bigl(\frac{c_1}{c_3}; q^{-h_2} z\Bigr) 
  \RF_{12}\Bigl(\frac{c_1}{c_2}; z\Bigr)
  \,,
\end{multline}
where $c_1$, $c_2$, $c_3$ are multiplicative spectral parameters and
$q^{-h} z$ is the same quantity as $\lambda - \gamma h$ in additive
notation.

In our discussions on Belavin's and Felder's R-matrices, we introduced
three kinds of backgrounds.  Let $n \in \Z$ be a charge that equals
$1$, $0$ and $-1$ in dark shaded, unshaded and light shaded
backgrounds, respectively.  We define two types of domain walls that
separate regions with different values of $n$.  Domain walls of one
type, which we draw with solid lines, have the property that $n$
increases by $1$ as they are crossed from the left to the right when
oriented upward.  Those of the other type, drawn with dotted lines,
change $n$ by $-1$ instead.  We only consider such configurations of
domain walls that $n$ stays in the range $\abs{n} \leq 1$.%
\footnote{The resulting tricolored surfaces are often represented by
  bipartite graphs whose nodes are placed on shaded regions and
  connected by edges that go through intersections of domain walls.}

Crossings of domain walls fall into four groups, distinguished by the
way in which the surrounding faces are shaded.  We assign a spectral
parameter to each domain wall, a dynamical variable to each unshaded
region, and the following Boltzmann weights to the crossings:
\begin{equation}
  \label{eq:MW}
  \begin{alignedat}{2}
  \begin{tikzpicture}[scale={1.2/sqrt(2)}, baseline=(x.base)]
    \node (x) at (0,0) {\vphantom{x}};
    
    \fill[ws] (-135:1) -- (-45:1) -- (45:1) -- (135:1) -- cycle;
    
    \fill[dshaded] (0,0) -- (45:1) -- (-45:1) -- cycle;
    \fill[lshaded] (0,0) -- (135:1) -- (-135:1) -- cycle;
    
    \draw[z->] (-135:1) node[below] {$a_1$} -- (45:1);
    \draw[z->] (-45:1) node[below] {$a_2$} -- (135:1);

    \node at (0,-{sqrt(2)/3}) {$z$};
    \node at (0,{sqrt(2)/3}) {$w$};
  \end{tikzpicture}
  &=
  M\Bigl(\frac{a_1}{a_2};z,w\Bigr)
  \,,
  &\qquad
  \begin{tikzpicture}[scale={1.2/sqrt(2)}, baseline=(x.base)]
    \node (x) at (0,0) {\vphantom{x}};
    
    \fill[ws] (-135:1) -- (-45:1) -- (45:1) -- (135:1) -- cycle;
    
    \fill[lshaded] (0,0) -- (45:1) -- (-45:1) -- cycle;
    \fill[dshaded] (0,0) -- (135:1) -- (-135:1) -- cycle;
    
    \draw[wz->] (-135:1) node[below] {$b_1$} -- (45:1);
    \draw[wz->] (-45:1) node[below] {$b_2$} -- (135:1);

    \node at (0,-{sqrt(2)/3}) {$z$};
    \node at (0,{sqrt(2)/3}) {$w$};
  \end{tikzpicture}
  &=
  \Mt\Bigl(\frac{b_1}{b_2};z,w\Bigr)
  \,,
  \\
  \begin{tikzpicture}[scale={1.2/sqrt(2)}, baseline=(x.base)]
    \node (x) at (0,0) {\vphantom{x}};
    
    \fill[ws] (-135:1) -- (-45:1) -- (45:1) -- (135:1) -- cycle;
    
    \fill[lshaded] (45:1) -- (0,0) -- (135:1) -- cycle;
    \fill[dshaded] (-45:1) -- (0,0) -- (-135:1) -- cycle;
    
    \draw[z->] (-135:1) node[below] {$a$} -- (45:1);
    \draw[wz->] (-45:1) node[below] {$b$} -- (135:1);

    \node at (-{sqrt(2)/3}, 0) {$z$};
    \node at ({sqrt(2)/3}, 0) {$w$};
  \end{tikzpicture}
  &=
  W\Bigl(\frac{a}{b};z,w\Bigr)
  \,,
  &
  \begin{tikzpicture}[scale={1.2/sqrt(2)}, baseline=(x.base)]
    \node (x) at (0,0) {\vphantom{x}};
    
    \fill[ws] (-135:1) -- (-45:1) -- (45:1) -- (135:1) -- cycle;
    
    \fill[dshaded] (45:1) -- (0,0) -- (135:1) -- cycle;
    \fill[lshaded] (-45:1) -- (0,0) -- (-135:1) -- cycle;
    
    \draw[wz->] (-135:1) node[below] {$b$} -- (45:1);
    \draw[z->] (-45:1) node[below] {$a$} -- (135:1);

    \node at (-{sqrt(2)/3}, 0) {$z$};
    \node at ({sqrt(2)/3}, 0) {$w$};
  \end{tikzpicture}
  &=
  \Wt\Bigl(\frac{b}{a};z,w\Bigr)
  \,.
  \end{alignedat}
\end{equation}
The functions used here are defined by
\begin{equation}
  \begin{alignedat}{2}
    M(a;z,w)
    &=
    \frac{\IB(1/a; z,w)}{\Gamma(a^{-N})}
    \,,
    &\qquad
    \Mt(a;z,w)
    &=
    M(a;w,z)
    \,,
    \\
    W(a;z,w)
    &=
    \IB(\sqrt{pq} a; z,w)
    \,,
    &
    \Wt(a;z,w)
    &=
    W(a;w,z)
    \,,
  \end{alignedat}
\end{equation}
with
\begin{equation}
  \IB(a; z,w)
  =
  \prod_{i,j} \Gamma\Bigl(a \frac{w_i}{z_j}\Bigr)
  \,.
\end{equation}
Note that these are symmetric meromorphic functions with respect to
each set of variables $(z_1, \dotsc, z_N)$ or $(w_1, \dotsc, w_N)$,
and therefore belong to $\V \otimes \V$.

Given a configuration of domain walls on a surface,%
\footnote{In addition to the constraint $\abs{n} \leq 1$, we require
  that for every bounded face, the domain walls bounding it all go
  upward locally for some choice of the vertical direction.
  See~\cite{Yagi:2016oum} for a discussion on this point.}
we define the associated partition function by the rule that the
dynamical variables assigned to the unshaded faces bounded by solid
and dotted lines are integrated over.  For each such variable~$z$, the
integration is performed with measure
\begin{equation}
  \label{eq:intIV}
  \int_{\T^{N-1}}
  \prod_{j=1}^{N-1} \frac{\rmd z_j}{2\pi\iu z_j}
  \IV(z)
  \,,
\end{equation}
where $\T$ is the unit circle with positive orientation in $\C$ and
\begin{equation}
  \IV(z)
  =
  \frac{(p;p)_\infty^{N-1} (q;q)_\infty^{N-1}}{N!}
  \prod_{i \neq j}
  \frac{1}{\Gamma(z_i/z_j)}
  \,.
\end{equation}
We do not integrate over the dynamical variables on those faces that
contain part of the boundary of the surface.  Thus, the partition
function is a meromorphic function of these variables.

The crossings of domain walls act on the lattice by concatenation, and
their action defines operators on the partition function.  The first
two crossings in \eqref{eq:MW} represent integral operators
$M(a_1/a_2)$, $\Mt(b_1/b_2) \in \End(\V)$ that act on $f \in \V$ as
\begin{align}
  \bigl(M(a) f\bigr)(w)
  &=
    \int_{\T^{N-1}}
    \prod_{j=1}^{N-1} \frac{\rmd z_j}{2\pi\iu z_j}
    \IV(z) M(a;z,w) f(z)
    \,,
  \\
  \bigl(\Mt(a) f\bigr)(w)
  &=
    \int_{\T^{N-1}}
    \prod_{k=1}^{N-1} \frac{\rmd z_j}{2\pi\iu z_j}
    \IV(z) \Mt(a;z,w) f(z)
    \,.
\end{align}
The remaining two crossings represent operators $W(a/b)$,
$\Wt(b/a) \in \End(\V \otimes \V)$ that act by multiplication:
\begin{align}
  \bigl(W(a) f\bigr)(z,w) 
  &=
    W(a;z,w) f(z,w)
    \,,
    \\
  \bigl(\Wt(a) f\bigr)(z,w) 
  &=
    \Wt(a;z,w) f(z,w)
  \,.
\end{align}
There is no integration here since concatenating these crossings do
not produce any unshaded bounded regions.  The operator $M$ was
introduced in \cite{MR2076912} for $N = 2$ and~\cite{MR2264067} for
$N \geq 2$.  Its action is known as the elliptic Fourier transform.

The unitarity relations hold for domain walls.  For example,
\begin{equation}
  \label{eq:unitarity1}
  \def\pathA{(0,0) to[out=0, in=180] (1,0.5) to[out=0, in=180] (2,0)}
  \def\pathB{(2,0.5) to[in=0,out=180] (1,0) to[in=0,out=180] (0,0.5)}
  \begin{tikzpicture}[baseline=(x.base)]
    \node (x) at (1,0.25) {\vphantom{x}};

    \begin{scope}
      \clip \pathA -- (2,-0.3) -- (0,-0.3) -- cycle;
      \fill[dshaded] (0,-0.3) rectangle (2,0.25);
    \end{scope}

    \begin{scope}
      \clip \pathB -- (0,0.8) -- (2,0.8) -- cycle;
      \fill[lshaded] (0,0.25) rectangle (2,0.8);
    \end{scope}

    \begin{scope}
      \clip \pathA -- \pathB -- cycle;
      \fill[ws] (0,0) rectangle (2,0.5);
    \end{scope}

    \draw[z->] node[left] {$a_1$} \pathA;
    \draw[z<-] \pathB node[left] {$a_2$}; 
  \end{tikzpicture}
  \ =
  \begin{tikzpicture}[baseline=(x.base)]
    \node (x) at (1,0.25) {\vphantom{x}};

    \fill[ws] (0,0) rectangle (2,0.5);

    \fill[dshaded] (0,-0.3) rectangle (2,0);
    \fill[lshaded] (0,0.5) rectangle (2,0.8);

    \draw[z->] (0,0) node[left] {$a_1$} -- (2,0);
    \draw[z->] (0,0.5) node[left] {$a_2$} -- (2,0.5);
  \end{tikzpicture}
  \iff
  M\Bigl(\frac{a_1}{a_2}\Bigr) M\Bigl(\frac{a_2}{a_1}\Bigr)
  =
  \id_\V
\end{equation}
is the inversion relation~\cite{MR2264067}
\begin{equation}
  \int_{\T^{N-1}}
  \prod_{j=1}^{N-1} \frac{\rmd x_j}{2\pi\iu x_j}
  \IV(x)
  M\Bigl(\frac{a_2}{a_1}; z,x\Bigr)
  M\Bigl(\frac{a_1}{a_2}; x,w\Bigr)
  =
  \delta(z,w)
  \,,
\end{equation}
where $\delta(z,w)$ is the delta function that equates the two sets of
dynamical variables~$z$ and~$w$ upon integration with respect to
measure~\eqref{eq:intIV}.  Another unitarity relation
\begin{equation}
  \label{eq:BB=0}
  \def\pathA{(0,0) to[out=0, in=180] (1,0.5) to[out=0, in=180] (2,0)}
  \def\pathB{(2,0.5) to[in=0,out=180] (1,0) to[in=0,out=180] (0,0.5)}
  \begin{tikzpicture}[baseline=(x.base)]
    \node (x) at (1,0.25) {\vphantom{x}};

    \fill[ws] (0,-0.3) rectangle (2,0.8);

    \path[name path=A] \pathA;
    \path[name path=B] \pathB;
    \path[name intersections={of=A and B, by={1,2}}];

    \begin{scope}
      \clip \pathA -- \pathB -- cycle;
      \fill[dshaded] (0,0) -- (0.5,0) -- (1) -- (0.5,0.5) -- (0,0.5) 
      -- cycle;
    \end{scope}

    \begin{scope}
      \clip \pathA -- \pathB -- cycle;
      \fill[dshaded]  (2,0) -- (1.5,0) -- (2) -- (1.5,0.5) -- (2,0.5) 
      -- cycle;
     \end{scope}

    \begin{scope}
      \clip \pathA -- \pathB -- cycle;
      \fill[lshaded] (0.5,0) -- (1.5,0) -- (2) -- (1.5,0.5) -- (0.5,0.5)
      -- (1) -- (0.5,0) -- cycle;
     \end{scope}

    \draw[wz->] node[left] {$b$} \pathA;
    \draw[z<-] \pathB node[left] {$a$};
  \end{tikzpicture}
  \ =
  \begin{tikzpicture}[baseline=(x.base)]
    \node (x) at (1,0.25) {\vphantom{x}};

    \fill[ws] (0,-0.3) rectangle (2,0.8);

    \fill[dshaded] (0,0) rectangle (2,0.5);

    \draw[wz->] (0,0) node[left] {$b$} -- (2,0);
    \draw[z->] (0,0.5) node[left] {$a$} -- (2,0.5);
  \end{tikzpicture}
  \iff
  \Wt\Bigl(\frac{b}{a}\Bigr) W\Bigl(\frac{a}{b}\Bigr)
  = \id_{\V \otimes \V}
\end{equation}
is a consequence of the identity $\Gamma(z) \Gamma(pq/z) = 1$.

One may hope that in the same vein, the Yang--Baxter equations for
domain walls can be established by means of some identities obeyed by
the elliptic gamma function.  This is not the case, unfortunately.
The reason is that regions with $\abs{n} > 1$ appear in these
equations, and the Boltzmann weights appropriate for crossings
involving such regions are not known (except in the case
$N = 2$~\cite{Bazhanov:2010kz, Maruyoshi:2016caf}).

The Bazhanov--Sergeev model avoids this difficulty by doubling the
number of lines.  Let us take a pair of solid and dotted lines, and
think of it as a single line.  We represent it by a double line:
\begin{equation}
  \label{eq:double-line}
  \begin{tikzpicture}[baseline=(x.base), scale=0.6]
    \node (x) at (1,0) {\vphantom{x}};

    \draw[dtr->] (0,0) node[left] {$(a,b)$} -- (2,0);
  \end{tikzpicture}
  \ = \
  \begin{tikzpicture}[baseline=(x.base), scale=0.6]
    \node (x) at (1,0) {\vphantom{x}};

    \draw[wz->] (0,0.25) node[left] {$b$} -- (2,0.25);
    \draw[z->] (0,-0.25) node[left] {$a$} -- (2,-0.25);
  \end{tikzpicture}
  \ .
\end{equation}
The BSDS R-operator
$\RBS((a_1,b_1), (a_2,b_2)) \in \End(\V \otimes \V)$ is the crossing
of two double lines in a dark shaded background:
\begin{equation}
  \label{eq:RBS-g}
  \RBS\bigl((a_1,b_1), (a_2,b_2)\bigr)
  =
  \begin{tikzpicture}[baseline=(x.base), scale=0.6]
    \node (x) at (1,1) {\vphantom{x}};

    \fill[dshaded] (0,0) rectangle (2,2);
    \draw[dtr->] (0,1) node[left] {$(a_1,b_1)$} -- (2,1);
    \draw[dtr->] (1,0) node[below] {$(a_2,b_2)$} -- (1,2);
  \end{tikzpicture}
  \ =
  \begin{tikzpicture}[baseline=(x.base), scale=0.6]
    \node (x) at (1,1) {\vphantom{x}};
    
    \fill[ws] (0,0) rectangle (2,2);

    \fill[dshaded] (0,0) rectangle (0.75,0.75);
    \fill[dshaded] (1.25,0) rectangle (2,0.75);
    \fill[dshaded] (0,1.25) rectangle (0.75,2);
    \fill[dshaded] (1.25,1.25) rectangle (2,2);
    \fill[lshaded] (0.75,0.75) rectangle (1.25,1.25);

    \draw[wz->] (0,1.25) node[left] {$b_1$}-- (2,1.25);
    \draw[z->] (0,0.75) node[left] {$a_1$} -- (2,0.75);
    \draw[wz->] (0.75,0) node[below] {$b_2$}-- (0.75,2);
    \draw[z->] (1.25,0) node[below] {$a_2$} -- (1.25,2);
  \end{tikzpicture}
  \ .
\end{equation}
In terms of the operators defined above, we have
\begin{equation}
  \label{eq:RBS}
  \RBS\bigl((a_1,b_1), (a_2,b_2)\bigr)
  =
  \P \Wt\Bigl(\frac{b_1}{a_2}\Bigr)
  M_2\Bigl(\frac{a_1}{a_2}\Bigr)
  \Mt_1\Bigl(\frac{b_1}{b_2}\Bigr)
  W\Bigl(\frac{a_1}{b_2}\Bigr)
 \,,
\end{equation}
where $\P$ swaps the two sets of variables, $(\P f)(z,w) = f(w,z)$.

The Bazhanov--Sergeev model is therefore a vertex model whose spins
are dynamical variables, and placed on a tricolor checkerboard lattice
such as the one shown in Fig.~\ref{fig:BT-D}.  Its Yang--Baxter
equation
\begin{equation}
  \begin{tikzpicture}[scale=0.6, baseline=(x.base)]
    \node (x) at (30:2) {\vphantom{x}};
    
    \fill[dshaded] (0,-0.5) rectangle ({3*sqrt(3)/2},2.5);

    \draw[dtr->] (0,2) -- ++(-30:3);
    \draw[dtr->] (0,0) -- ++(30:3);
    \draw[dtr->] (-30:1) -- ++(0,3);
  \end{tikzpicture}
  \ = \
  \begin{tikzpicture}[scale=0.6, baseline=(x.base)]
    \node (x) at (30:1) {\vphantom{x}};
    
    \fill[dshaded] (0,-1) rectangle ({3*sqrt(3)/2},2);

    \draw[dtr->] (0,1) -- ++(-30:3);
    \draw[dtr->] (0,0) -- ++(30:3);
    \draw[dtr->] (-30:2) -- ++(0,3);
  \end{tikzpicture}
\end{equation}
does not involve any undesirable regions, and can be derived from an
integral identity for the elliptic gamma function~\cite{MR1846786,
  MR2044635, MR2630038}.

Of course, using a pair of solid and dotted lines we can make another
line, which we represent by a thick solid line:
\begin{equation}
  \label{eq:thick-line}
  \begin{tikzpicture}[baseline=(x.base), scale=0.6]
    \node (x) at (1,0) {\vphantom{x}};

    \draw[tr->] (0,0) node[left] {$(a,b)$} -- (2,0);
  \end{tikzpicture}
  \ = \
  \begin{tikzpicture}[baseline=(x.base), scale=0.6]
    \node (x) at (1,0) {\vphantom{x}};

    \draw[z->] (0,0.25) node[left] {$a$} -- (2,0.25);
    \draw[wz->] (0,-0.25) node[left] {$b$} -- (2,-0.25);
  \end{tikzpicture}
  \ .
\end{equation}
The R-operator defined by the crossing of two thick lines in a
light shaded background,
\begin{equation}
  \label{eq:RBSt-g}
  \RBSt\bigl((a_1,b_1), (a_2,b_2)\bigr)
  =
  \begin{tikzpicture}[baseline=(x.base), scale=0.6]
    \node (x) at (1,1) {\vphantom{x}};

    \fill[lshaded] (0,0) rectangle (2,2);
    \draw[tr->] (0,1) node[left] {$(a_1,b_1)$} -- (2,1);
    \draw[tr->] (1,0) node[below] {$(a_2,b_2)$} -- (1,2);
  \end{tikzpicture}
  \ =
  \begin{tikzpicture}[baseline=(x.base), scale=0.6]
    \node (x) at (1,1) {\vphantom{x}};
    
    \fill[ws] (0,0) rectangle (2,2);

    \fill[lshaded] (0,0) rectangle (0.75,0.75);
    \fill[lshaded] (1.25,0) rectangle (2,0.75);
    \fill[lshaded] (0,1.25) rectangle (0.75,2);
    \fill[lshaded] (1.25,1.25) rectangle (2,2);
    \fill[dshaded] (0.75,0.75) rectangle (1.25,1.25);

    \draw[z->] (0,1.25) node[left] {$a_1$}-- (2,1.25);
    \draw[wz->] (0,0.75) node[left] {$b_1$} -- (2,0.75);
    \draw[z->] (0.75,0) node[below] {$a_2$}-- (0.75,2);
    \draw[wz->] (1.25,0) node[below] {$\,b_2$} -- (1.25,2);
  \end{tikzpicture}
  \ ,
\end{equation}
is an equally good solution of the Yang--Baxter equation.

\section{Unification}
\label{sec:unification}

In the previous section we introduced dashed lines in shaded and
unshaded backgrounds to describe the Belavin and Jimbo--Miwa--Okado
models, and solid and dotted lines representing domain walls to
formulate the Bazhanov--Sergeev model.  In fact, the three kinds of
lines can coexist in a single model without spoiling integrability.
In this section we construct this integrable lattice model that
unifies the three models.

\subsection{Intertwining operators}

The unified model, if exists, should allow dashed lines to cross solid
and dotted lines.  Conversely, in order to establish the existence of
the unified model, we just need to determine what should be assigned
to such crossings and show that they are compatible with
integrability.

Consider crossings that involve dark shaded and unshaded regions.
They define two matrix-valued functions $S(a;z)$,
$S'(a;z) \in \End(V) \otimes \V$:
\begin{equation}
  \begin{tikzpicture}[scale=0.6, baseline=(x.base)]
    \node (x) at (1,1) {\vphantom{x}};

    \fill[ws] (0,0) rectangle (1,2);
    \fill[dshaded] (1,0) rectangle (2,2);

    \draw[dr->] (0,1) node[left]{$c$} -- (2,1);
    \draw[z->] (1,0) node[below] {$a$} -- (1,2);

    \node at (0.5,1.5) {$z$};
  \end{tikzpicture}
  \
  =
  S\Bigl(\frac{c}{a};z\Bigr)
  \,,
  \qquad
  \begin{tikzpicture}[scale=0.6, baseline=(x.base)]
    \node (x) at (1,1) {\vphantom{x}};

    \fill[dshaded] (0,0) rectangle (1,2);
    \fill[ws] (1,0) rectangle (2,2);

    \draw[dr->] (0,1) node[left] {$c$} -- (2,1);;
    \draw[wz->] (1,0) node[below] {$b$} -- (1,2);

    \node at (1.5,0.5) {$z$};
  \end{tikzpicture}
  \ =
  S'\Bigl(\frac{c}{b};z\Bigr)
  \,.
\end{equation}
Similarly, those involving light shaded and unshaded regions define
matrix-valued functions $\St(a;z)$,
$\St'(a;z) \in \End(V) \otimes \V$:
\begin{equation}
  \begin{tikzpicture}[scale=0.6, baseline=(x.base)]
    \node (x) at (1,1) {\vphantom{x}};

    \fill[ws] (0,0) rectangle (1,2);
    \fill[lshaded] (1,0) rectangle (2,2);

    \draw[dr->] (0,1) node[left]{$c$} -- (2,1);
    \draw[wz->] (1,0) node[below] {$b$} -- (1,2);

    \node at (0.5,1.5) {$z$};
  \end{tikzpicture}
  \ =
  \St\Bigl(\frac{c}{b};z\Bigr)
  \,,
  \qquad
  \begin{tikzpicture}[scale=0.6, baseline=(x.base)]
    \node (x) at (1,1) {\vphantom{x}};

    \fill[lshaded] (0,0) rectangle (1,2);
    \fill[ws] (1,0) rectangle (2,2);

    \draw[dr->] (0,1) node[left] {$c$} -- (2,1);;
    \draw[z->] (1,0) node[below] {$a$} -- (1,2);

    \node at (1.5,0.5) {$z$};
  \end{tikzpicture}
  \ =
  \St'\Bigl(\frac{c}{a};z\Bigr)
  \,.
\end{equation}
We assume that these functions depend on the spectral parameters of
the relevant lines only through their ratio.  This is a reasonable
assumption since $\RB$, $\RF$ and $\RBS$ all have this property.

A dashed line segment supporting a state $e_i \in V$ changes the
dynamical variable $\lambda$ by $-\gamma\omega_i$ according to the
rule shown in \eqref{eq:lambda-change}.  As such, it acts on the
partition function as the inverse of the difference operator
\begin{equation}
  \label{eq:Ti}
  \Delta_i\colon \lambda \mapsto \lambda + \gamma\omega_i
  \,.
\end{equation}
The above crossings thus represent the matrices of difference
operators $S(c/a)$, $S'(c/b)$, $\St(c/b)$,
$\St'(c/a) \in \End(V \otimes \V)$ whose elements are given by
\begin{alignat}{2}
  S(a)^j_i
  &=
  S(a;z)^j_i \Delta_i^{-1}
  \,,
  &\qquad
  S'(a)^j_i
  &=
  \Delta_j^{-1} S'(a;z)^j_i
  \,,
  \\
  \St(a)^j_i
  &=
  \St(a;z)^j_i \Delta_i^{-1}
  \,,
  &
  \St'(a)^j_i
  &=
  \Delta_j^{-1} \St'(a;z)^j_i
  \,.
\end{alignat}
As we will see, $S$ and $S'$ intertwine $\RB$ and $\RF$, whereas $\St$
and $\St'$ intertwine $\RBt$ and $\RF$.  The unified model should be
constructed from these intertwining operators, together with
the R-matrices~\eqref{eq:RB-g}, \eqref{eq:RBt-g} and~\eqref{eq:RF-g}
as well as the operators~\eqref{eq:MW}.

\subsection{Yang--Baxter equations with one dashed line}
\label{sec:YBEw1dl}

Let us determine the intertwining operators.  To this end, we analyze
Yang--Baxter equations that involve a single dashed line.  Part of
this analysis was essentially done in~\cite{Sergeev:1992ap,
  Quano:1992wc}.  A similar analysis was carried out for $N = 2$
in~\cite{Derkachov:2012iv}.

First, we look at the following Yang--Baxter equation that contains
$S$ and $\St'$:%
\footnote{By $S(a) \St'(b)$ we mean an $\End(\V)$-valued matrix
  whose $(i,j)$ component is
  $\sum_k S(a; z)^i_k \Delta_k^{-1} \St'(b; z)^k_j$.}
\begin{equation}
  \begin{tikzpicture}[scale=0.6, baseline=(x.base)]
    \node (x) at (1,{sqrt(3)*3/4}) {\vphantom{x}};

    \fill[dshaded] (1,{sqrt(3)}) -- ++(60:1) -- ++(1,0) -- (2.5,0)
    -- (2,0) -- cycle;

    \fill[lshaded] (1,{sqrt(3)}) -- ++(120:1) -- ++(-1,0) -- (-0.5,0)
    -- (0,0) -- cycle;

    \fill[ws] (1,{sqrt(3)}) -- ++(60:1) -- ++(-1,0) -- cycle;
    \fill[ws] (1,{sqrt(3)}) -- ++(-60:2) -- ++(-2,0) -- cycle;

    \draw[z->] (0,0) node[below] {$a_1$} -- ++(60:3);
    \draw[z->] (2,0) node[below] {$a_2$} -- ++(120:3);
    \draw[dr->] (-0.5,{sqrt(3)/2}) node[left] {$c$} -- ++(3,0);
  \end{tikzpicture}
  \ =
  \begin{tikzpicture}[scale=0.6, baseline=(x.base)]
    \node (x) at (0.5,{sqrt(3)*3/4}) {\vphantom{x}};

    \fill[dshaded] (0.5,{sqrt(3)/2}) -- ++(60:2) -- ++(0.5,0) -- (2,0)
    -- (1,0) -- cycle;

    \fill[lshaded] (0.5,{sqrt(3)/2}) -- ++(120:2) -- ++(-0.5,0) --
    (-1,0) -- (0,0) -- cycle;

    \fill[ws] (0.5,{sqrt(3)/2}) -- ++(60:2) -- ++(-2,0) -- cycle;
    \fill[ws] (0.5,{sqrt(3)/2}) -- ++(-60:1) -- ++(-1,0) -- cycle;

    \draw[z->] (0,0) node[below] {$a_1$} -- ++(60:3);
    \draw[z->] (1,0) node[below] {$a_2$} -- ++(120:3);
    \draw[dr->] (-1,{sqrt(3)}) node[left] {$c$} -- ++(3,0);
  \end{tikzpicture}
  \iff
  \begin{aligned}
    &
    M\Bigl(\frac{a_1}{a_2}\Bigr)
    S\Bigl(\frac{c}{a_2}\Bigr)
    \St'\Bigl(\frac{c}{a_1}\Bigr)
    \\
    &\qquad
    =
    S\Bigl(\frac{c}{a_1}\Bigr)
    \St'\Bigl(\frac{c}{a_2}\Bigr)
    M\Bigl(\frac{a_1}{a_2}\Bigr)
    \,.
  \end{aligned}
\end{equation}
This is a relation between integral operators on $V \otimes \V$.  In
components, it reads
\begin{multline}
  \label{eq:int-MSS}
  \int_{\T^{N-1}}
  \prod_{l=1}^{N-1} \frac{\rmd x_l}{2\pi\iu x_l}
  \IV(x)
  \IB\Bigl(\frac{a_2}{a_1}; x,w\Bigr)
  \sum_{i,k}
  S\Bigl(\frac{c}{a_2}; x\Bigr)^j_k
  \St'\Bigl(\frac{c}{a_1}; \Delta_k^{-1} x\Bigr)^k_i
  f^i(\Delta_k^{-1} x)
  \\
  =
  \sum_{i,k}
  S\Bigl(\frac{c}{a_1}; w\Bigr)^j_k
  \St'\Bigl(\frac{c}{a_2}; \Delta_k^{-1} w\Bigr)^k_i
  \int_{\T^{N-1}}
  \prod_{l=1}^{N-1} \frac{\rmd z_l}{2\pi\iu z_l}
  \IV(z)
  \IB\Bigl(\frac{a_2}{a_1}; z, \Delta_k^{-1} w\Bigr)
  f^i(z)
  \,,
\end{multline}
where $f \in V \otimes \V$ is a $V$-valued function on which the
operators act.

For each summand on the left-hand side, let us change the integration
variables from $x$ to $z = \Delta_k^{-1} x$ so that the same factor
$f^i(z)$ appears in the integrands of both sides.  After this change
of variables, the integration is performed with the same measure
$\prod_{l=1}^{N-1} \rmd z_l/2\pi\iu z_l$ as the one used on the
right-hand side, but over a different contour given by
$\abs{\Delta_k z_l} = 1$.  In order to compare the two sides, we
deform this contour to the one given by $\abs{z_l} = 1$ or
$\abs{\Delta_k z_l} = \abs{q}^{\delta_{kl} - 1/N}$.  For this
deformation to leave the integral unchanged, the integrand should have
no simple poles in the domain sandwiched by the two contours, for
all~$k$ and~$l$.%
\footnote{In principle, it is possible that the contributions from the
  simple poles in this domain cancel out.  We do not consider this
  possibility since we want to keep $f$ as general as possible.}

The poles of the factor $\IB(a_2/a_1; \Delta_k z,w)$ are located at
$\Delta_k z_i = p^m q^n a_2 w_j/a_1$, where $i$, $j = 1$, $\dotsc$,
$N$ and $m$, $n$ are nonnegative integers.  None of these poles enters
the relevant domain if and only if there exist no pairs $(m,n)$ such
that
$\abs{q}^{1 - 1/N} \leq \abs{p^m q^n a_2 w_j/a_1} \leq \abs{q}^{-1/N}$
for some $j$.  This is the case if and only if
\begin{equation}
  \label{eq:a2a1w}
  \Bigabs{\frac{a_2}{a_1} w_j}  < \abs{q}^{1 - 1/N}
\end{equation}
for all $j$.  For the moment we assume that the remaining part of the
integrand does not introduce harmful poles either.

Comparing the integrands, we see that for the above equation to hold
for any $f \in V \otimes \V$ satisfying this assumption, $S(a;z)$ and
$\St'(a;z)$ should satisfy
\begin{multline}
  \sum_k
  \IV(\Delta_k z)
  \IB\Bigl(\frac{a_2}{a_1}; \Delta_k z,w\Bigr)
  S\Bigl(\frac{c}{a_2}; \Delta_k z\Bigr)^j_k
  \St'\Bigl(\frac{c}{a_1}; z\Bigr)^k_i
  \\
  =
  \sum_k
  S\Bigl(\frac{c}{a_1}; w\Bigr)^j_k
  \St'\Bigl(\frac{c}{a_2}; \Delta_k^{-1} w\Bigr)^k_i
  \IV(z)
  \IB\Bigl(\frac{a_2}{a_1}; z,\Delta_k^{-1} w\Bigr)
  \,.
\end{multline}
It follows that we should have
\begin{multline}
  \IV(\Delta_i z)
  \IB\Bigl(\frac{a_2}{a_1}; \Delta_i z, \Delta_j w\Bigr)
  \biggl[S\Bigl(\frac{c}{a_1}; \Delta_j w\Bigr)^{-1}
  S\Bigl(\frac{c}{a_2}; \Delta_i z\Bigr)\biggr]^j_i
  \\
  =
  \biggl[\St'\Bigl(\frac{c}{a_2}; w\Bigr)
  \St'\Bigl(\frac{c}{a_1}; z\Bigr)^{-1}\biggr]^j_i
  \IV(z)
  \IB\Bigl(\frac{a_2}{a_1}; z,w\Bigr)
  \,.
\end{multline}
Graphically, this equation can be expressed as
\begin{equation}
  \begin{tikzpicture}[scale=0.6, baseline=(x.base)]
    \node (x) at (1,{sqrt(3)*3/4}) {\vphantom{x}};

    \fill[ws] (1,{sqrt(3)}) -- ++(60:1) -- ++(1,0) -- (2.5,0)
    -- (2,0) -- cycle;

    \fill[ws] (1,{sqrt(3)}) -- ++(120:1) -- ++(-1,0) -- (-0.5,0)
    -- (0,0) -- cycle;

    \fill[lshaded] (1,{sqrt(3)}) -- ++(60:1) -- ++(-1,0) -- cycle;
    \fill[dshaded] (1,{sqrt(3)}) -- ++(-60:2) -- ++(-2,0) -- cycle;

    \draw[z->] (0,0) node[below] {$a_2$} -- ++(60:3);
    \draw[z<-] (2,0) node[below] {$a_1$} -- ++(120:3);
    \draw[dr->] (-0.5,{sqrt(3)/2}) node[left] {$c$} -- ++(3,0);

    \node (O) at (1,{sqrt(3)*4/6}) {};
    \node at ($(O) + (210:{sqrt(3)*9/12})$) {$z$};
    \node at ($(O) + (-30:{sqrt(3)*9/12})$) {$w$};
  \end{tikzpicture}
  \ =
  \begin{tikzpicture}[scale=0.6, baseline=(x.base)]
    \node (x) at (0.5,{sqrt(3)*3/4}) {\vphantom{x}};

    \fill[ws] (0.5,{sqrt(3)/2}) -- ++(60:2) -- ++(0.5,0) -- (2,0)
    -- (1,0) -- cycle;

    \fill[ws] (0.5,{sqrt(3)/2}) -- ++(120:2) -- ++(-0.5,0) --
    (-1,0) -- (0,0) -- cycle;

    \fill[lshaded] (0.5,{sqrt(3)/2}) -- ++(60:2) -- ++(-2,0) -- cycle;
    \fill[dshaded] (0.5,{sqrt(3)/2}) -- ++(-60:1) -- ++(-1,0) -- cycle;

    \draw[z->] (0,0) node[below] {$a_2$} -- ++(60:3);
    \draw[z<-] (1,0) node[below] {$a_1$} -- ++(120:3);
    \draw[dr->] (-1,{sqrt(3)}) node[left] {$c$} -- ++(3,0);

    \node (O) at (0.5,{sqrt(3)*5/6}) {};
    \node at ($(O) + (-150:{sqrt(3)*7/12})$) {$z$};
    \node at ($(O) + (-30:{sqrt(3)*7/12})$) {$w$};
  \end{tikzpicture}
  \ .
\end{equation}
Canceling the elliptic gamma functions that appear on the two sides,
we get
\begin{equation}
  \label{eq:St'St'/SS}
  \frac{\bigl[S(c/a_1; \Delta_j w)^{-1} S(c/a_2; \Delta_i z)\bigr]^j_i}
       {\bigl[\St'(c/a_2;w) \St'(c/a_1;z)^{-1}\bigr]^j_i}
  =
  \frac{\prod_{l (\neq j)}
        \theta(q^{-1} a_2 w_l/a_1 z_i)}
       {\prod_{k (\neq i)}
        \theta(a_2 w_j/a_1 z_k)}
         \prod_{k (\neq i)}
  \frac{\theta(z_i/z_k)}
       {\theta(q^{-1} z_k/z_i)}
  \,.
\end{equation}

To solve the above equation, we use the matrices $\Phi(u,\lambda)$ and
$\Psi(u,\lambda)$ introduced in section~\ref{sec:VFC} in the context
of the vertex--face correspondence.  We have~\cite{MR1463830}
\begin{equation}
  \bigl[\Phi(v, \mu)^{-1} \Phi(u, \lambda)\bigr]^j_i
  =
  \frac{\theta_1\bigl(v + (u-v)/N + \mu_j - \lambda_i\bigr)}
       {\theta_1(v)}
  \prod_{l (\neq j)}
  \frac{\theta_1\bigl((u-v)/N + \mu_l - \lambda_i\bigr)}
       {\theta_1(\mu_l - \mu_j)}
  \,.
\end{equation}
Thus, if we set
\begin{align}
  S(a;z)
  &=
  a^{-N/2} \Psi(u,\lambda)
  \,,
  \\
  \St'(a;z)
  &=
  a^{-N/2} Z^{N/2} \Phi(u,\lambda)^T
  \,,
\end{align}
with $a = e^{2\pi\iu u/N}$ and $Z = \diag(z_1, \dotsc, z_N)$, then
they satisfy
\begin{align}
  \label{eq:S-1S}
  \bigl[S(b;w)^{-1} S(a;z)\bigr]^j_i
  &=
    \frac{\theta(b^{-N} b w_j/a z_i)}
         {\theta(b^{-N})}
    \frac{\prod_{l (\neq j)} \theta(b w_l/a z_i)}
         {\prod_{k (\neq i)} \theta(z_k/z_i)}
  \,,
  \\
  \label{eq:StSt-1}
  \bigl[\St'(a;w) \St'(b;z)^{-1}\bigr]^j_i
  &=
    \frac{\theta(b^{-N} b w_j/a z_i)}
         {\theta(b^{-N})}
    \prod_{k (\neq i)} 
    \frac{\theta(b w_j/a z_k)}
         {\theta(z_i/z_k)}
\end{align}
and solve \eqref{eq:St'St'/SS}.  We have chosen the normalization
factor $a^{-N/2}$ so as to remove factors of $a/b$ from these
formulas.

Let us check the validity of the contour deformation for this
solution.  Apart from the function $f^i(z)$, the potentially dangerous
factors in the integrand on the left-hand side of~\eqref{eq:int-MSS}
are $\IV(T_k z)$ as well as $1/\prod_{l (\neq k)} \theta(qz_k/z_l)$
contained in $S(c/a_2; \Delta_k z)^j_k$.  Their product consists of
factors of the form $\Gamma(x)/\Gamma(x/q)$ or
$1/\Gamma(x) \Gamma(1/x)$ with $x = z_i/z_j$ for some $i$, $j$, but
neither of these functions has poles except at $x = 0$.  Thus, the
contour deformation is justified as long as inequality
\eqref{eq:a2a1w} is satisfied and $f(z)$ has no simple poles in the
domain swept out by the deformation.

Next, we analyze the Yang--Baxter equation
\begin{equation}
  \label{eq:MtStS'}
  \begin{tikzpicture}[scale=0.6, baseline=(x.base)]
    \node (x) at (1,{sqrt(3)*3/4}) {\vphantom{x}};

    \fill[lshaded] (1,{sqrt(3)}) -- ++(60:1) -- ++(1,0) -- (2.5,0)
    -- (2,0) -- cycle;

    \fill[dshaded] (1,{sqrt(3)}) -- ++(120:1) -- ++(-1,0) -- (-0.5,0)
    -- (0,0) -- cycle;

    \fill[ws] (1,{sqrt(3)}) -- ++(60:1) -- ++(-1,0) -- cycle;
    \fill[ws] (1,{sqrt(3)}) -- ++(-60:2) -- ++(-2,0) -- cycle;

    \draw[wz->] (0,0) node[below] {$b_1$} -- ++(60:3);
    \draw[wz->] (2,0) node[below] {$b_2$} -- ++(120:3);
    \draw[dr->] (-0.5,{sqrt(3)/2}) node[left] {$c$} -- ++(3,0);
  \end{tikzpicture}
  \ =
  \begin{tikzpicture}[scale=0.6, baseline=(x.base)]
    \node (x) at (0.5,{sqrt(3)*3/4}) {\vphantom{x}};

    \fill[lshaded] (0.5,{sqrt(3)/2}) -- ++(60:2) -- ++(0.5,0) -- (2,0)
    -- (1,0) -- cycle;

    \fill[dshaded] (0.5,{sqrt(3)/2}) -- ++(120:2) -- ++(-0.5,0) --
    (-1,0) -- (0,0) -- cycle;

    \fill[ws] (0.5,{sqrt(3)/2}) -- ++(60:2) -- ++(-2,0) -- cycle;
    \fill[ws] (0.5,{sqrt(3)/2}) -- ++(-60:1) -- ++(-1,0) -- cycle;

    \draw[wz->] (0,0) node[below] {$b_1$} -- ++(60:3);
    \draw[wz->] (1,0) node[below] {$b_2$} -- ++(120:3);
    \draw[dr->] (-1,{sqrt(3)}) node[left] {$c$} -- ++(3,0);
  \end{tikzpicture}
  \iff
  \begin{aligned}
    &
    \Mt\Bigl(\frac{b_1}{b_2}\Bigr)
    \St\Bigl(\frac{c}{b_2}\Bigr)
    S'\Bigl(\frac{c}{b_1}\Bigr)
    \\
    &\qquad
    =
    \St\Bigl(\frac{c}{b_1}\Bigr)
    S'\Bigl(\frac{c}{b_2}\Bigr)
    \Mt\Bigl(\frac{b_1}{b_2}\Bigr)
    \,.
  \end{aligned}
\end{equation}
The same contour deformation argument as above leads to the condition
\begin{multline}
  \IV(\Delta_i z)
  \IB\Bigl(\frac{b_2}{b_1}; \Delta_j w, \Delta_i z\Bigr)
  \biggl[\St\Bigl(\frac{c}{b_1}; \Delta_j w\Bigr)^{-1}
  \St\Bigl(\frac{c}{b_2}; \Delta_i z\Bigr)\biggr]^j_i
  \\
  =
  \biggl[S'\Bigl(\frac{c}{b_2}; w\Bigr)
  S'\Bigl(\frac{c}{b_1}; z\Bigr)^{-1}\biggr]^j_i 
  \IV(z)
  \IB\Bigl(\frac{b_2}{b_1}; w,z\Bigr)
  \,,
\end{multline}
or graphically,
\begin{equation}
  \begin{tikzpicture}[scale=0.6, baseline=(x.base)]
    \node (x) at (1,{sqrt(3)*3/4}) {\vphantom{x}};
  
    \fill[ws] (1,{sqrt(3)}) -- ++(60:1) -- ++(1,0) -- (2.5,0)
    -- (2,0) -- cycle;
  
    \fill[ws] (1,{sqrt(3)}) -- ++(120:1) -- ++(-1,0) -- (-0.5,0)
    -- (0,0) -- cycle;
  
    \fill[dshaded] (1,{sqrt(3)}) -- ++(60:1) -- ++(-1,0) -- cycle;
    \fill[lshaded] (1,{sqrt(3)}) -- ++(-60:2) -- ++(-2,0) -- cycle;
  
    \draw[wz->] (0,0) node[below] {$b_2$} -- ++(60:3);
    \draw[wz<-] (2,0) node[below] {$b_1$} -- ++(120:3);
    \draw[dr->] (-0.5,{sqrt(3)/2}) node[left] {$c$} -- ++(3,0);
  
    \node (O) at (1,{sqrt(3)*4/6}) {};
    \node at ($(O) + (210:{sqrt(3)*9/12})$) {$z$};
    \node at ($(O) + (-30:{sqrt(3)*9/12})$) {$w$};
  \end{tikzpicture}
  \ =
  \begin{tikzpicture}[scale=0.6, baseline=(x.base)]
    \node (x) at (0.5,{sqrt(3)*3/4}) {\vphantom{x}};
  
    \fill[ws] (0.5,{sqrt(3)/2}) -- ++(60:2) -- ++(0.5,0) -- (2,0)
    -- (1,0) -- cycle;
  
    \fill[ws] (0.5,{sqrt(3)/2}) -- ++(120:2) -- ++(-0.5,0) --
    (-1,0) -- (0,0) -- cycle;
  
    \fill[dshaded] (0.5,{sqrt(3)/2}) -- ++(60:2) -- ++(-2,0) -- cycle;
    \fill[lshaded] (0.5,{sqrt(3)/2}) -- ++(-60:1) -- ++(-1,0) -- cycle;
  
    \draw[wz->] (0,0) node[below] {$b_2$} -- ++(60:3);
    \draw[wz<-] (1,0) node[below] {$b_1$} -- ++(120:3);
    \draw[dr->] (-1,{sqrt(3)}) node[left] {$c$} -- ++(3,0);
  
    \node (O) at (0.5,{sqrt(3)*5/6}) {};
    \node at ($(O) + (-150:{sqrt(3)*7/12})$) {$z$};
    \node at ($(O) + (-30:{sqrt(3)*7/12})$) {$w$};
  \end{tikzpicture}
  \ .
\end{equation}
From this equation we get
\begin{equation}
  \label{eq:S'S'/StSt}
  \begin{split}
    \frac{\bigl[\St(c/b_1; \Delta_j w)^{-1}
           \St(c/b_2; \Delta_i z)\bigr]^j_i}
         {\bigl[S'(c/b_2;w) S'(c/b_1;z)^{-1}\bigr]^j_i}
    &=
    \frac{\prod_{k (\neq i)} \theta(q^{-1} b_2 z_k/b_1 w_j)}
         {\prod_{l (\neq j)} \theta(b_2 z_i/b_1 w_l)}
    \prod_{k (\neq i)}
    \frac{\theta(z_i/z_k)}{\theta(q^{-1} z_k/z_i)}
  \\
  &=       
  \Bigl(\frac{z_i}{w_j}\Bigr)^N
  \frac{\bigl[\St'(tc/b_1; \Delta_j w)
         \St'(tc/b_2; \Delta_i z)^{-1}\bigr]^j_i}
       {\bigl[S(tc/b_2;w)^{-1} S(tc/b_1;z)\bigr]^j_i}
  \,,
  \end{split}
\end{equation}
where $t \in \C$ is an arbitrary parameter.

Finally, the Yang--Baxter equation
\begin{equation}
  \begin{tikzpicture}[scale=0.6, baseline=(x.base)]
    \node (x) at (1,{sqrt(3)*3/4}) {\vphantom{x}};

    \fill[ws] (1,{sqrt(3)}) -- ++(60:1) -- ++(1,0) -- (2.5,0)
    -- (2,0) -- cycle;

    \fill[ws] (1,{sqrt(3)}) -- ++(120:1) -- ++(-1,0) -- (-0.5,0)
    -- (0,0) -- cycle;

    \fill[lshaded] (1,{sqrt(3)}) -- ++(60:1) -- ++(-1,0) -- cycle;
    \fill[dshaded] (1,{sqrt(3)}) -- ++(-60:2) -- ++(-2,0) -- cycle;

    \draw[z->] (0,0) node[below] {$a$} -- ++(60:3);
    \draw[wz->] (2,0) node[below] {$b$} -- ++(120:3);
    \draw[dr->] (-0.5,{sqrt(3)/2}) node[left] {$c$} -- ++(3,0);
  \end{tikzpicture}
  \ = \
  \begin{tikzpicture}[scale=0.6, baseline=(x.base)]
    \node (x) at (0.5,{sqrt(3)*3/4}) {\vphantom{x}};

    \fill[ws] (0.5,{sqrt(3)/2}) -- ++(60:2) -- ++(0.5,0) -- (2,0)
    -- (1,0) -- cycle;

    \fill[ws] (0.5,{sqrt(3)/2}) -- ++(120:2) -- ++(-0.5,0) --
    (-1,0) -- (0,0) -- cycle;

    \fill[lshaded] (0.5,{sqrt(3)/2}) -- ++(60:2) -- ++(-2,0) -- cycle;
    \fill[dshaded] (0.5,{sqrt(3)/2}) -- ++(-60:1) -- ++(-1,0) -- cycle;

    \draw[z->] (0,0) node[below] {$a$} -- ++(60:3);
    \draw[wz->] (1,0) node[below] {$b$} -- ++(120:3);
    \draw[dr->] (-1,{sqrt(3)}) node[left] {$c$} -- ++(3,0);
  \end{tikzpicture}
  \iff
  \begin{aligned}
    &
    W\Bigl(\frac{a}{b}\Bigr)
    S'\Bigl(\frac{c}{b}\Bigr)
    S\Bigl(\frac{c}{a}\Bigr)
    \\
    &\qquad
    =
    \St'\Bigl(\frac{c}{a}\Bigr)
    \St\Bigl(\frac{c}{b}\Bigr)
    W\Bigl(\frac{a}{b}\Bigr)
  \end{aligned}
\end{equation}
gives the relation
\begin{multline}
  \IB\Bigl(\sqrt{pq} \frac{a}{b}; z, \Delta_j w\Bigr)
  \biggl[S'\Bigl(\frac{c}{b}; w\Bigr)
  S\Bigl(\frac{c}{a};z\Bigr)\biggr]^j_i
  \\
  =
  \biggl[\St'\Bigl(\frac{c}{a};w\Bigr)
  \St\Bigl(\frac{c}{b};z\Bigr)\biggr]^j_i
  \IB\Bigl(\sqrt{pq} \frac{a}{b}; \Delta_i^{-1} z, w\Bigr) 
  \,.
\end{multline}
We can rewrite it as
\begin{equation}
  \label{eq:StSt'/S'S}
  \begin{split}
    \frac{\bigl[\St'(c/a;w) \St(c/b;z)\bigr]^j_i}
         {\bigl[S'(c/b;w) S(c/a;z)\bigr]^j_i}
    &=
    \frac{
      \prod_{k (\neq i)}
      \theta(q^{-1/N} \sqrt{pq} a w_j/b z_k)
    }{
      \prod_{l (\neq j)}
      \theta(q^{-1/N} \sqrt{pq} a w_l/b z_i)
    }
    \\
    &=
    (-1)^{N-1}
    \frac{\bigl[\St'(c/a;w) \St'(q^{-1/N} \sqrt{pq} c/b;z)^{-1} Z^N\bigr]^j_i}
    {\bigl[S(q^{-1/N} \sqrt{pq} c/b;w)^{-1} S(c/a;z)\bigr]^j_i}
  \,.
  \end{split}
\end{equation}
The two equations \eqref{eq:S'S'/StSt} and \eqref{eq:StSt'/S'S} are
solved by
\begin{align}
  S'(a;z)
  &= S(q^{-1/N} \sqrt{pq} a; z)^{-1}
  \,,
  \\
  \St(a;z)
  &=
  (-1)^{N-1} \St'(q^{-1/N} \sqrt{pq} a;z)^{-1} Z^N
  \,.
\end{align}
The remaining Yang--Baxter equations with one dashed line follow from
those considered above.

The contour deformation argument used in the analysis of relation
\eqref{eq:MtStS'} is valid for the above solution, provided that
\begin{equation}
  \Bigabs{\frac{b_1}{b_2} w_j} > \abs{q}^{-1/N}
\end{equation}
for all $j$ and the two sides of the relation act on functions that
have no simple poles in the deformation domain.  This inequality
ensures that $\IB(b_2/b_1; w, \Delta_k z)$ has no simple poles in the
domain.  To see that the other relevant factors have no simple poles
there either, we calculate
$\sum_j \St'(a; w)^l_j \IV(\Delta_k z) \St(c/b_2; \Delta_k z)^j_k$ and
$\sum_i S'(c/b_1; z)^k_i S(a; w)^i_l$ using formulas~\eqref{eq:S-1S}
and \eqref{eq:StSt-1}.  Recalling that
$\IV(\Delta_k z)/ \prod_{l (\neq k)} \theta(qz_k/z_l)$ has no poles in
the domain under consideration, we find that these quantities do not
have harmful poles in $z$.  Hence, the same is true for
$\IV(\Delta_k z) \St(c/b_2; \Delta_k z)^j_k$ and $S'(c/b_1; z)^k_i$.

\subsection{Yang--Baxter equations with two dashed lines}

Now we move on to the Yang--Baxter equations that contain two dashed
lines.  Those with a solid line are
\begin{equation}
  \label{eq:RSS}
  \begin{tikzpicture}[scale=0.6, baseline=(x.base)]
    \node (x) at (30:2) {\vphantom{x}};

    \fill[ws] (-30:1) rectangle ++({-sqrt(3)/2},3);
    \fill[dshaded] (-30:1) rectangle ++({sqrt(3)},3);

    \draw[dr->] (0,0) node[left] {$c_2$} -- ++(30:3);
    \draw[dr->] (0,2) node[left] {$c_1$} -- ++(-30:3);
    \draw[z->] (-30:1) node[below] {$a$} -- ++(0,3);

    \node (O) at ({sqrt(3)*4/6},1) {};
    \node at ($(O) + (120:{sqrt(3)*9/12})$) {$z$};
  \end{tikzpicture}
  \ =
  \begin{tikzpicture}[scale=0.6, baseline=(x.base)]
    \node (x) at (30:1) {\vphantom{x}};

    \fill[ws] (-30:2) rectangle ++({-sqrt(3)},3);
    \fill[dshaded] (-30:2) rectangle ++({sqrt(3)/2},3);

    \draw[dr->] (0,0) node[left] {$c_2$} -- ++(30:3);
    \draw[dr->] (0,1) node[left] {$c_1$} -- ++(-30:3);
    \draw[z->] (-30:2) node[below] {$a$} -- ++(0,3);

    \node (O) at ({sqrt(3)*5/6},0.5) {};
    \node at ($(O) + (120:{sqrt(3)*7/12})$) {$z$};
  \end{tikzpicture}
  \iff
  \begin{aligned}
    & \RB\Bigl(\frac{c_1}{c_2}\Bigr)
    S_1\Bigl(\frac{c_1}{a};z\Bigr)
    S_2\Bigl(\frac{c_2}{a}; q^{-h_1} z\Bigr)
    \\
    & \qquad
    =
    S_2\Bigl(\frac{c_2}{a};z\Bigr)
    S_1\Bigl(\frac{c_1}{a}; q^{-h_2} z\Bigr)
    \RF\Bigl(\frac{c_1}{c_2}; z\Bigr)
  \end{aligned}
\end{equation}
and
\begin{equation}
  \label{eq:RStSt}
  \begin{tikzpicture}[scale=0.6, baseline=(x.base)]
    \node (x) at (30:2) {\vphantom{x}};

    \fill[lshaded] (-30:1) rectangle ++({-sqrt(3)/2},3);
    \fill[ws] (-30:1) rectangle ++({sqrt(3)},3);

    \draw[dr->] (0,0) node[left] {$c_2$} -- ++(30:3);
    \draw[dr->] (0,2) node[left] {$c_1$} -- ++(-30:3);
    \draw[z->] (-30:1) node[below] {$a$} -- ++(0,3);

    \node (O) at ({sqrt(3)*4/6},1) {};
    \node at ($(O) + (-60:{sqrt(3)*7/12})$) {$z$};
  \end{tikzpicture}
  \ =
  \begin{tikzpicture}[scale=0.6, baseline=(x.base)]
    \node (x) at (30:1) {\vphantom{x}};

    \fill[lshaded] (-30:2) rectangle ++({-sqrt(3)},3);
    \fill[ws] (-30:2) rectangle ++({sqrt(3)/2},3);

    \draw[dr->] (0,0) node[left] {$c_2$} -- ++(30:3);
    \draw[dr->] (0,1) node[left] {$c_1$} -- ++(-30:3);
    \draw[z->] (-30:2) node[below] {$a$} -- ++(0,3);

    \node (O) at ({sqrt(3)*5/6},0.5) {};
    \node at ($(O) + (-60:{sqrt(3)*9/12})$) {$z$};
  \end{tikzpicture}
  \iff
  \begin{aligned}
    &\RF\Bigl(\frac{c_1}{c_2}; z\Bigr)
    \St'_1\Bigl(\frac{c_1}{a}; q^{h_2} z\Bigr)
    \St'_2\Bigl(\frac{c_2}{a}; z\Bigr)
    \\
    &\qquad
    =
    \St'_2\Bigl(\frac{c_2}{a}; q^{h_1} z\Bigr)
    \St'_1\Bigl(\frac{c_1}{a}; z\Bigr)
    \RBt\Bigl(\frac{c_1}{c_2}\Bigr)
    \,.
  \end{aligned}
\end{equation}
These equations describe the vertex--face correspondence discussed in
section~\ref{sec:VFC}.  The first one is relation \eqref{eq:RBPsiPsi},
while the second is a consequence of relation \eqref{eq:RBPhiPhi} and
the fact that $\RF(a,z)^{kl}_{ij} = 0$ unless $\{i,j\} = \{k,l\}$.

We can replace the solid lines above with dotted ones and obtain two
more relations.  Let us look at the Yang--Baxter equation
\begin{equation}
  \begin{tikzpicture}[scale=0.6, baseline=(x.base)]
    \node (x) at (30:2) {\vphantom{x}};

    \fill[ws] (-30:1) rectangle ++({-sqrt(3)/2},3);
    \fill[lshaded] (-30:1) rectangle ++({sqrt(3)},3);

    \draw[dr->] (0,0) node[left] {$c_2$} -- ++(30:3);
    \draw[dr->] (0,2) node[left] {$c_1$} -- ++(-30:3);
    \draw[wz->] (-30:1) node[below] {$b$} -- ++(0,3);

    \node (O) at ({sqrt(3)*4/6},1) {};
    \node at ($(O) + (120:{sqrt(3)*9/12})$) {$z$};
  \end{tikzpicture}
  \ =
  \begin{tikzpicture}[scale=0.6, baseline=(x.base)]
    \node (x) at (30:1) {\vphantom{x}};

    \fill[ws] (-30:2) rectangle ++({-sqrt(3)},3);
    \fill[lshaded] (-30:2) rectangle ++({sqrt(3)/2},3);

    \draw[dr->] (0,0) node[left] {$c_2$} -- ++(30:3);
    \draw[dr->] (0,1) node[left] {$c_1$} -- ++(-30:3);
    \draw[wz->] (-30:2) node[below] {$b$} -- ++(0,3);

    \node (O) at ({sqrt(3)*5/6},0.5) {};
    \node at ($(O) + (120:{sqrt(3)*7/12})$) {$z$};
  \end{tikzpicture}
  \iff
  \begin{aligned}
    &
    \RBt\Bigl(\frac{c_1}{c_2}\Bigr)
    \St_1\Bigl(\frac{c_1}{b}; z\Bigr)
    \St_2\Bigl(\frac{c_2}{b}; q^{-h_1} z\Bigr)
    \\
    &\qquad
    =
    \St_2\Bigl(\frac{c_2}{b}; z\Bigr)
    \St_1\Bigl(\frac{c_1}{b}; q^{-h_2} z\Bigr)
    \RF\Bigl(\frac{c_1}{c_2}; z\Bigr)
    \,.
  \end{aligned}
\end{equation}
It can be rewritten as
\begin{multline}
  \sum_{m,n}
  \St'\Bigl(q^{-1/N} \sqrt{pq} \frac{c_1}{b}; \Delta_l^{-1} z\Bigr)^k_m
  \St'\Bigl(q^{-1/N} \sqrt{pq} \frac{c_2}{b}; z\Bigr)^l_n
  \RBt\Bigl(\frac{c_1}{c_2}\Bigr)^{mn}_{ij}
  \\
  =
  \sum_{m,n}
  \RF\Bigl(\frac{c_1}{c_2}; z\Bigr)^{kl}_{mn}
  \St'\Bigl(q^{-1/N} \sqrt{pq} \frac{c_2}{b}; \Delta_m^{-1} z\Bigr)^n_j
  \St'\Bigl(q^{-1/N} \sqrt{pq} \frac{c_1}{b}; z\Bigr)^m_i
  \,.
\end{multline}
Replacing $z$ with $\Delta_k \Delta_l z$ and using the identity
\begin{equation}
  \label{eq:St'=St'}
  \St'(a; \Delta_j z)^j_i
  =
  \St'(q^{-1+1/N} a; z)^j_i
  \,,
\end{equation}
we see that this equation reduces to the intertwining relation
\eqref{eq:RStSt}.

The last Yang--Baxter equation with two dashed lines,
\begin{equation}
  \label{eq:RS'S'}
  \begin{tikzpicture}[scale=0.6, baseline=(x.base)]
    \node (x) at (30:2) {\vphantom{x}};

    \fill[dshaded] (-30:1) rectangle ++({-sqrt(3)/2},3);
    \fill[ws] (-30:1) rectangle ++({sqrt(3)},3);

    \draw[dr->] (0,0) node[left] {$c_2$} -- ++(30:3);
    \draw[dr->] (0,2) node[left] {$c_1$} -- ++(-30:3);
    \draw[wz->] (-30:1) node[below] {$a$} -- ++(0,3);

    \node (O) at ({sqrt(3)*4/6},1) {};
    \node at ($(O) + (-60:{sqrt(3)*7/12})$) {$z$};
  \end{tikzpicture}
  \ =
  \begin{tikzpicture}[scale=0.6, baseline=(x.base)]
    \node (x) at (30:1) {\vphantom{x}};

    \fill[dshaded] (-30:2) rectangle ++({-sqrt(3)},3);
    \fill[ws] (-30:2) rectangle ++({sqrt(3)/2},3);

    \draw[dr->] (0,0) node[left] {$c_2$} -- ++(30:3);
    \draw[dr->] (0,1) node[left] {$c_1$} -- ++(-30:3);
    \draw[wz->] (-30:2) node[below] {$a$} -- ++(0,3);

    \node (O) at ({sqrt(3)*5/6},0.5) {};
    \node at ($(O) + (-60:{sqrt(3)*9/12})$) {$z$};
  \end{tikzpicture}
  \iff
  \begin{aligned}
    &\RF\Bigl(\frac{c_1}{c_2}; z\Bigr)
    S'_1\Bigl(\frac{c_1}{a}; q^{h_2} z\Bigr)
    S'_2\Bigl(\frac{c_2}{a}; z\Bigr)
    \\
    &\qquad
    =
    S'_2\Bigl(\frac{c_2}{a}; q^{h_1} z\Bigr)
    S'_1\Bigl(\frac{c_1}{a}; z\Bigr)
    \RB\Bigl(\frac{c_1}{c_2}\Bigr)
    \,,
  \end{aligned}
\end{equation}
can be verified in a similar manner.  An easier way is to make use of
the relation
\begin{equation}
  \begin{tikzpicture}[scale=0.6, baseline=(x.base)]
    \node (x) at (0,1) {\vphantom{x}};

    \fill[ws] (0,0) rectangle (2,2);
    \fill[dshaded] (0.75,2) rectangle (1.25,0);

    \draw[dr->] (0,1) node[left] {$c$} -- (2,1);
    \draw[z->] (0.75,0) node[below] {$a$}-- (0.75,2);
    \draw[wz->]   (1.25,0) node[below] {$b$} -- (1.25,2);
  \end{tikzpicture}
  \ = \
  \frac{\theta\bigl((b'/c)^N\bigr)}{\theta\bigl(q(b'/c)^N\bigr)}
  \times
  \begin{tikzpicture}[scale=0.6, baseline=(x.base)]
    \node (x) at (0,1) {\vphantom{x}};

    \fill[ws] (0,0) rectangle (2,2);
    \fill[dshaded] (0.75,2) rectangle (1.25,0);

    \draw[dr->] (0,1) node[left] {$c$} -- (2,1);
    \draw[z->] (0.75,0) node[below] {$a$}-- (0.75,2);
    \draw[z<-]   (1.25,0) node[below] {$b'$} -- (1.25,2);
  \end{tikzpicture}
  \ ;
  \quad
  b' = \frac{b}{\sqrt{pq}}
  \,,
\end{equation}
which follows from the identity
$[S(a; T_j z)^{-1}]^j_i = [S(q^{-1/N} a;z)^{-1}]^j_i \theta(q
a^{-N})/\theta(a^{-N})$.  This relation and the intertwining relation
\eqref{eq:RSS} imply
\begin{equation}
  \begin{tikzpicture}[scale=0.6, baseline=(x.base)]
    \node (x) at (30:2) {\vphantom{x}};

    \fill[ws] (0,-0.5) rectangle ++({sqrt(3)*3/2},3);
    \fill[dshaded, shift={(-0.25,0)}] (-30:1) -- ++(0,3) -- ++(0.5,0)
    -- ++(0,-3) -- cycle;

    \draw[dr->] (0,0) -- ++(30:3);
    \draw[dr->] (0,2) -- ++(-30:3);
    \draw[z->, shift={(-0.25,0)}] (-30:1) -- ++(0,3);
    \draw[wz->, shift={(0.25,0)}] (-30:1) -- ++(0,3);
  \end{tikzpicture}
  \ = \
  \begin{tikzpicture}[scale=0.6, baseline=(x.base)]
    \node (x) at (30:1) {\vphantom{x}};

    \fill[ws] (0,-1) rectangle ++({sqrt(3)*3/2},3);
    \fill[dshaded, shift={(-0.25,0)}] (-30:2) -- ++(0,3) -- ++(0.5,0)
    -- ++(0,-3) -- cycle;

    \draw[dr->] (0,0) -- ++(30:3);
    \draw[dr->] (0,1) -- ++(-30:3);
    \draw[z->, shift={(-0.25,0)}] (-30:2) -- ++(0,3);
    \draw[wz->, shift={(0.25,0)}] (-30:2) -- ++(0,3);
  \end{tikzpicture}
  \ .
\end{equation}
Moving the solid line on the right-hand side past the crossing of the
dashed lines, we deduce that relation~\eqref{eq:RS'S'} holds.

\subsection{L-operators and elliptic quantum groups}

In our framework, an L-operator is constructed from two intertwining
operators composed in $V$ in a way consistent with shading.  The
Yang--Baxter equations with one dashed line studied in
section~\ref{sec:YBEw1dl} may be thought of as describing the action
of the operators $M$, $\Mt$ and $W$ on L-operators.

An important L-operator $\LB(a,b) \in \End(V \otimes \V)$ is given by
the crossing of a dashed line and a double line~\eqref{eq:double-line}
in a dark shaded background:
\begin{equation}
  \label{eq:LB}
  \LB\Bigl(\frac{c}{a},\frac{c}{b}\Bigr)
  =
  \begin{tikzpicture}[scale=0.6, baseline=(x.base)]
    \node (x) at (1,1) {\vphantom{x}};

    \fill[dshaded] (0,0) rectangle (2,2);
    \draw[dtr->] (1,0) node[below] {$(a,b)$} -- (1,2);
    \draw[dr->] (0,1) node[left] {$c$} -- (2,1);
  \end{tikzpicture}
  \ =
  \begin{tikzpicture}[scale=0.6, baseline=(x.base)]
    \node (x) at (1,1) {\vphantom{x}};

    \fill[ws] (0,0) rectangle (2,2);

    \fill[dshaded] (0,0) rectangle (0.75,2);
    \fill[dshaded] (1.25,0) rectangle (2,2);
    \draw[dr->] (0,1) node[left] {$c$} -- (2,1);
    \draw[wz->] (0.75,0) node[below] {$b$} -- (0.75,2);
    \draw[z->] (1.25,0) node[below] {$a$} -- (1.25,2);
  \end{tikzpicture}
  \ =
  S\Bigl(\frac{c}{a}\Bigr) S'\Bigl(\frac{c}{b}\Bigr) 
  \,.
\end{equation}
This L-operator was constructed in~\cite{Sergeev:1992ap, Quano:1992wc,
  MR1190749}.  It satisfies two Yang--Baxter equations,
\begin{equation}
  \begin{tikzpicture}[scale=0.6, baseline=(x.base)]
    \node (x) at (30:2) {\vphantom{x}};
    
    \fill[dshaded] (0,-0.5) rectangle ({3*sqrt(3)/2},2.5);

    \draw[dtr->] (0,0) -- ++(30:3);
    \draw[dtr->] (-30:1) -- ++(0,3);
    \draw[dr->] (0,2) -- ++(-30:3);
  \end{tikzpicture}
  \ = \
  \begin{tikzpicture}[scale=0.6, baseline=(x.base)]
    \node (x) at (30:1) {\vphantom{x}};
    
    \fill[dshaded] (0,-1) rectangle ({3*sqrt(3)/2},2);

    \draw[dtr->] (0,0) -- ++(30:3);
    \draw[dtr->] (-30:2) -- ++(0,3);
    \draw[dr->] (0,1) -- ++(-30:3);
  \end{tikzpicture}
  \iff
  \LB_1 \LB_2 \RBS
  =
  \RBS \LB_2 \LB_1
\end{equation}
and
\begin{equation}
  \begin{tikzpicture}[scale=0.6, baseline=(x.base)]
    \node (x) at (30:2) {\vphantom{x}};
    
    \fill[dshaded] (0,-0.5) rectangle ({3*sqrt(3)/2},2.5);

    \draw[dtr->] (-30:1) -- ++(0,3);
    \draw[dr->] (0,0) -- ++(30:3);
    \draw[dr->] (0,2) -- ++(-30:3);
  \end{tikzpicture}
  \ = \
  \begin{tikzpicture}[scale=0.6, baseline=(x.base)]
    \node (x) at (30:1) {\vphantom{x}};
    
    \fill[dshaded] (0,-1) rectangle ({3*sqrt(3)/2},2);

    \draw[dtr->] (-30:2) -- ++(0,3);
    \draw[dr->] (0,0) -- ++(30:3);
    \draw[dr->] (0,1) -- ++(-30:3);
  \end{tikzpicture}
  \iff
  \RB
  \LB_1 
  \LB_2
  =
  \LB_2
  \LB_1 
  \RB
  \,.
\end{equation}
By definition, an L-operator for an R-matrix is an operator that
satisfies an RLL relation of this kind together with that R-matrix.
We see that $\LB$ is an L-operator for both Belavin's R-matrix and the
BSDS R-operator.

In a certain trigonometric limit, the Bazhanov--Sergeev model for
$N = 2$ reduces to the chiral Potts model~\cite{Bazhanov:2010kz}.  On
the other hand, the Belavin model for $N = 2$ is the eight-vertex
model, which becomes the six-vertex model in this limit.  Therefore,
what we have just found is an elliptic counterpart of the fact that a
single L-operator satisfies two RLL relations, one for the chiral
Potts model and another one for the six-vertex
model~\cite{Bazhanov:1989nc}.  For $N > 2$, our result should
generalize the relation established in~\cite{Bazhanov:1990qk}.

Placing the crossing of a dashed line and a thick
line~\eqref{eq:thick-line} in an unshaded background, we obtain
another L-operator $\LF(a, b) \in \End(V \otimes \V \otimes \V)$.  Let
$z$ and $w$ be the dynamical variables for the first and second
factors of $\V$, respectively, and denote by $T_{z,i}$ the difference
operator~\eqref{eq:Ti} acting on functions of $z$.  Then, $\LF(a, b)$
is a matrix of difference operators with elements
\begin{equation}
  \label{eq:LF}
  \LF(a, b)^j_i
  =
  \Delta_{w,j}^{-1} \LF(a, b; z,w)^j_i \Delta_{z,i}^{-1}
  \,,
\end{equation}
where the matrix-valued function $\LF(a,b; z,w)$ is given by
\begin{equation}
  \label{eq:LF-g}
  \LF\Bigl(\frac{c}{a},\frac{c}{b}; z,w\Bigr)
  =
  \begin{tikzpicture}[scale=0.6, baseline=(x.base)]
    \node (x) at (1,1) {\vphantom{x}};

    \fill[ws] (0,0) rectangle (2,2);
    \draw[dr->] (0,1) node[left] {$c$} -- (2,1);
    \draw[tr->] (1,0) node[below] {$(a,b)$} -- (1,2);

    \node at (0.5,1.5) {$z$};
    \node at (1.5,0.5) {$w$};
  \end{tikzpicture}
  \ =
  \begin{tikzpicture}[scale=0.6, baseline=(x.base)]
    \node (x) at (0,1) {\vphantom{x}};

    \fill[ws] (0,0) rectangle (0.75,2);
    \fill[ws] (1.25,0) rectangle (2,2);
    \fill[dshaded] (0.75,2) rectangle (1.25,0);

    \draw[dr->] (0,1) node[left] {$c$} -- (2,1);
    \draw[z->] (0.75,0) node[below] {$a$}-- (0.75,2);
    \draw[wz->] (1.25,0) node[below] {$b$} -- (1.25,2);

    \node at (0.325,1.5) {$z$};
    \node at (1.625,0.5) {$w$};
  \end{tikzpicture}
  \ =
  S'\Bigl(\frac{c}{b}; w\Bigr) S\Bigl(\frac{c}{a}; z\Bigr)
  \,.
\end{equation}
This L-operator satisfies an RLL relation with Felder's R-matrix:
\begin{equation} 
  \begin{tikzpicture}[scale=0.6, baseline=(x.base)]
    \node (x) at (30:2) {\vphantom{x}};
    
    \fill[ws] (0,-0.5) rectangle ({3*sqrt(3)/2},2.5);

    \draw[dr->] (0,0) node[left] {$c_2$} -- ++(30:3);
    \draw[dr->] (0,2) node[left] {$c_1$} -- ++(-30:3);
    \draw[tr->] (-30:1) node[below] {$(a,b)$} -- ++(0,3);

    \node (O) at ({sqrt(3)*4/6},1) {};
    \node at ($(O) + (120:{sqrt(3)*9/12})$) {$z$};
    \node at ($(O) + (-60:{sqrt(3)*7/12})$) {$w$};
  \end{tikzpicture}
  \ =
  \begin{tikzpicture}[scale=0.6, baseline=(x.base)]
    \node (x) at (30:1) {\vphantom{x}};
    
    \fill[ws] (0,-1) rectangle ({3*sqrt(3)/2},2);

    \draw[dr->] (0,0) node[left] {$c_2$} -- ++(30:3);
    \draw[dr->] (0,1) node[left] {$c_1$} -- ++(-30:3);
    \draw[tr->] (-30:2) node[below] {$(a,b)$} -- ++(0,3);

    \node (O) at ({sqrt(3)*5/6},0.5) {};
    \node at ($(O) + (120:{sqrt(3)*7/12})$) {$z$};
    \node at ($(O) + (-60:{sqrt(3)*9/12})$) {$w$};
  \end{tikzpicture}
  \ ,
\end{equation}
or more explicitly,
\begin{multline}
  \label{eq:RLL-F}
  \RF\Bigl(\frac{c_1}{c_2}; w\Bigr)
  \LF_1\Bigl(\frac{c_1}{a}, \frac{c_1}{b}; z, q^{h_2} w\Bigr)
  \LF_2\Bigl(\frac{c_2}{a}, \frac{c_2}{b}; q^{-h_1} z, w\Bigr)
  \\
  =
  \LF_2\Bigl(\frac{c_2}{a}, \frac{c_2}{b}; z, q^{h_1} w\Bigr)
  \LF_1\Bigl(\frac{c_1}{a}, \frac{c_1}{b}; q^{-h_2} z, w\Bigr)
  \RF\Bigl(\frac{c_1}{c_2}; z\Bigr)
  \,.
\end{multline}
Note that $\LF$ is obtained from $\LB$ by an interchange of the
intertwining operators in the product.

The last relation leads to the notion of the elliptic quantum
group~$E_{\tau, \gamma/2}(\slf_N)$~\cite{Felder:1994pb,
  Felder:1994be}.  Set $L(u) = \LF(c/a, c/b)$ with
$c = e^{2\pi\iu u/N}$, and let $\lambda$, $\mu$ be additive dynamical
variables corresponding to $z$, $w$, respectively.  Then, $L(u)$ act
on functions $f(\lambda)$, $g(\mu)$ by
\begin{equation}
  \label{eq:Lf=fL}
  L(u)^j_i f(\lambda)
  = f(\lambda - \gamma\omega_i) L(u)^j_i \,,
  \qquad
  L(u)^j_i g(\mu)
  = g(\mu - \gamma\omega_j) L(u)^j_i
\end{equation}
and satisfies
\begin{equation}
  \label{eq:RLL-EQG}
  \sum_{k,l}
  \RF(u_1 - u_2, \mu)^{mn}_{kl}
  L(u_1)^k_i
  L(u_2)^l_j
  =
  \sum_{k,l}
  \RF(u_1 - u_2, \lambda)^{kl}_{ij}
  L(u_2)^n_l
  L(u_1)^m_k
  \,.
\end{equation}
As an algebra, $E_{\tau, \gamma/2}(\slf_N)$ is generated by
meromorphic functions of $\lambda$ and those of $\mu$, and the matrix
elements of an L-operator $L(u)$ and its inverse $L^{-1}(u)$, obeying
these relations.  It is further endowed with the structure of an
$\hf$-Hopf algebroid~\cite{MR1645196, MR2045681, MR2539704}.

In fact, $\LF$ gives a representation of $E_{\tau, \gamma/2}(\slf_N)$.
Let us introduce an $\hf$-module
\begin{equation}
  W = \bigoplus_{\mu \in \mu_0 + \Lambda} \C w_\mu
  \,,
\end{equation}
where $\mu_0 \in \hf^*$ is a parameter, $\Lambda$ is the root lattice
of $\slf_N$, and $w_\mu$ is a generator of weight~$\mu$.  If we define
$L(u,\lambda) \in \End(V \otimes W)$ by
\begin{equation}
  L(u,\lambda) (e_i \otimes w_\mu)
  =
  \sum_j
  L^F\Bigl(\frac{c}{a}, \frac{c}{b}; z, q^{-\mu - \omega_i} z\Bigr)^j_i
  e_j \otimes w_{\mu + \omega_i - \omega_j}
  \,,
\end{equation}
then the RLL relation~\eqref{eq:RLL-F} implies
\begin{multline}
  \RF_{12}(u_1 - u_2, \lambda - \gamma h_3)
  L_{13}(u_1,\lambda)
  L_{23}(u_2, \lambda - \gamma h_1)
  \\
  =
  L_{23}(u_2, \lambda)
  L_{13}(u_1, \lambda - \gamma h_2)
  \RF_{12}(u_1 - u_2, \lambda)
  \,.
\end{multline}
A diagonalizable $\hf$-module $W$ and an $\End(V \otimes W)$-valued
meromorphic function $L(u,\lambda)$ that commutes with the action of
$\hf$ and obeys this relation constitute a representation of
$E_{\tau, \gamma/2}(\slf_N)$~\cite{Felder:1994pb, Felder:1994be}.

\section{Surface defects and elliptic quantum groups}
\label{sec:SD}

The proposal of this paper is that the integrable lattice model
formulated in the previous section can be realized by intersecting
branes in string theory.  In turn, this proposal implies that a family
of surface defects in four-dimensional $\CN = 1$ supersymmetric field
theories correspond to transfer matrices constructed from L-operators
for elliptic quantum groups and variants thereof.  In this section we
discuss the brane construction and this correspondence.  We provide
evidence for the proposal by computing some classes of transfer
matrices and showing that they reproduce known gauge theory results.

\subsection{Brane construction}

A brane construction of the Bazhanov--Sergeev model was found
in~\cite{Yamazaki:2012cp}, and a closely related construction of the
eight-vertex model was proposed in~\cite{Maruyoshi:2016caf};
see~\cite{Yagi:2016oum} for a review.  The brane construction of the
integrable lattice model in question is a natural extension of these
constructions.

Consider a stack of $N$ coincident D5-branes supported on
$\R^{3,1} \times \Sigma \times 0$ in type IIB string theory in
spacetime $\R^{3,1} \times T^*\Sigma \times \R^2$, where $\Sigma$ is a
compact surface embedded in the cotangent bundle $T^*\Sigma$ as the
zero section.  The surface $\Sigma$ wrapped by the D5-branes becomes
an unshaded background surface on which a lattice is placed.

\begin{figure}
  \centering
  \begin{tikzpicture}[scale=0.6, align at top]
    \draw[ws, shift={(0.16,0.16)}] (0,0) rectangle (3,2);
    \draw[ws, shift={(0.08,0.08)}] (0,0) rectangle (3,2);
    \draw[ws] (0,0) rectangle (3,2);
    \node[left] at (0,1.65) {D5};
    \node[left] at (-0.6,0.5) {D3};

    \draw[fill=red!10] (0,1) .. controls (0.75,1.5) and (1.75,0.5) .. (2.965,1) --
    (2.3,0.3) .. controls (0.85,-0.2) and (0.05,0.8) .. (-0.7,0.3) -- cycle;

    \draw[densely dotted] (0,1.05) -- (0,0.45);
  \end{tikzpicture}
  \quad
  \begin{tikzpicture}[scale=0.6, align at top]
    \draw[draw=none, shift={(0.16,0.16)}] (0,0) rectangle (3,2);
    \fill[ws] (0,0) rectangle (3,2);
    \draw[dr->] (0,1) .. controls (0.75,1.5) and (1.75,0.5) .. (3,1);

    \draw[frame] (0,0) rectangle (3,2);
  \end{tikzpicture}

  \caption{A D3-brane ending on the D5-branes creates a dashed line.}
  \label{fig:D3-D5}
\end{figure}
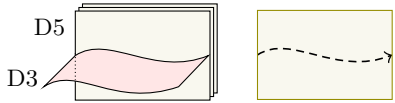

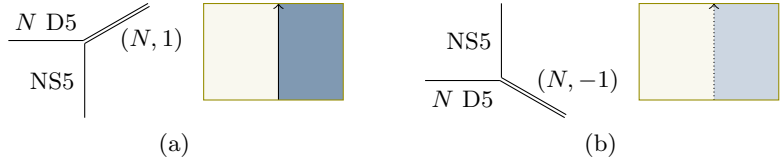
\begin{figure}
  \centering
  \subfloat[\label{fig:5BW-a}]{
    \begin{tikzpicture}[align at top]
      \begin{scope}[shift={(-0.02,0.04)}]
        \draw (0,0) -- node[above] {$N$ D5} (1,0);
        \draw (1,0) -- ++(30:1);
      \end{scope}

      \draw (1,-1) -- node[left] {NS5} (1,0);
      \draw (1,0) -- node[below right=-4pt] {$(N,1)$} ++(30:1);
    \end{tikzpicture}
    \begin{tikzpicture}[align at top]
      \fill[ws] (0,0) rectangle ({1+sqrt(3)/2},1.3);
      \fill[dshaded] (1,0) rectangle ({1+sqrt(3)/2},1.3);
      \draw[z->] (1,0) -- (1,1.3);
      \draw[frame] (0,0) rectangle ({1+sqrt(3)/2},1.3);
    \end{tikzpicture}
  }
  \qquad
  \subfloat[\label{fig:5BW-b}]{
    \begin{tikzpicture}[align at top]
      \draw (0,0) -- node[below] {$N$ D5} (1,0);
      \draw (1,0) -- ++(-30:1);

      \begin{scope}[shift={(0.02,0.04)}]
        \draw (1,0) -- node[left] {NS5} (1,1);
        \draw (1,0) -- node[above right=-4pt] {$(N,-1)$} ++(-30:1);
      \end{scope}
    \end{tikzpicture}
    \begin{tikzpicture}[align at top]
      \fill[ws] (0,0) rectangle ({1+sqrt(3)/2},1.3);
      \fill[lshaded] (1,0) rectangle ({1+sqrt(3)/2},1.3);
      \draw[wz->] (1,0) -- (1,1.3);
      \draw[frame] (0,0) rectangle ({1+sqrt(3)/2},1.3);
    \end{tikzpicture}
  }
  \caption{An NS5-brane combines with $N$ D5-branes to form an $(N,1)$
    or $(N,-1)$ 5-brane.}
  \label{fig:5BW}
\end{figure}

To draw a lattice on $\Sigma$, we introduce D3-branes and NS5-branes.
Take a D3-brane that ends on the D5-branes along a curve $C$ on
$\Sigma$ and is supported on $\R^{1,1} \times \Sigma_C \times 0$,
where $\Sigma_C$ is a suitably chosen surface in $T^*\Sigma$ such that
$\Sigma_C \cap \Sigma = C$.  The curve $C$ represents a dashed line
(Fig.~\ref{fig:D3-D5}).  Similarly, NS5-branes intersecting the
D5-branes create solid and dotted lines.  The difference from the
D3-brane case is that an NS5-brane cannot simply terminate on the
D5-branes.  Rather, these 5-branes merge into a bound state, either an
$(N,1)$ 5-brane (Fig.~\ref{fig:5BW-a}) or an $(N,-1)$ 5-brane
(Fig.~\ref{fig:5BW-b}), depending on the relative positions of the
5-branes.  Therefore, the curve along which the NS5-brane and the
D5-branes meet is a domain wall on $\Sigma$, which separates a
D5-brane region and an $(N, \pm 1)$ 5-brane region.  In our graphical
notation, $(N,1)$ and $(N,-1)$ 5-brane regions are indicated by dark
and light shading, respectively.

After making a lattice on $\Sigma$ with D3- and NS5-branes, we perform
a Wick rotation and compactify $\R^4$ to the quotient $\CM_{p,q}$ of
$\C^2 \setminus (0,0)$ by the equivalence relation
$(z_1,z_2) \sim (pz_1,qz_2)$.  We put the D3-branes on the submanifold
$\{z_2 = 0\}$ of~$\CM_{p,q}$, which is an elliptic curve $E_p$ with
modulus $p = e^{2\pi\iu\tau}$.  Computing the partition function of
the brane system in this new spacetime, we obtain a supersymmetric
index~\cite{Romelsberger:2005eg, Kinney:2005ej, Festuccia:2011ws,
  Closset:2013vra}.  It is identified with the partition function of
the integrable lattice model defined in the previous section, placed
on this lattice.

The supersymmetric index is protected against continuous changes of
various parameters of the theory.  In the present case, these
parameters include the shapes of the D3- and NS5-branes.  Hence, the
partition function of the lattice model is invariant under continuous
deformations of lattice lines; it is a topological invariant of the
lattice.

The Yang--Baxter equation states that the partition function remains
unchanged as one line moves past the intersection of two other lines.
This is a stronger statement than the topological invariance, which
does not rule out the possibility of a phase transition occurring when
the three lines meet at a point.  The absence of such a phase
transition follows from string dualities.  Topologically, $\CM_{p,q}$
is $S^1 \times S^3$.  If we apply T-duality along $S^1$, we obtain a
dual brane system in type IIA string theory, which we can further lift
to M-theory.  These duality operations turn the D5-branes into
M5-branes supported on $S^3 \times \Sigma \times S^1$, where the last
$S^1$ is the M-theory circle, the 11th dimension that emerged in the
process.  On the other hand, the D3- and NS5-branes are mapped to M2-
and M5-branes, respectively.  Since they are generically supported at
separate points on the M-theory circle, the seemingly singular
situation of three lines meeting at a point is rendered nonsingular in
the M-theory picture.

The emergent extra dimension does not only ensure that the
Yang--Baxter equation holds, but also provides spectral
parameters~\cite{Costello:2013zra, Costello:2013sla}: the coordinates
of the M2- and M5-branes on the M-theory circle.  Together, the
Yang--Baxter equation and the presence of spectral parameters imply
the integrability of the lattice model.

In the original frame, the spectral parameters are the holonomies of
the $\U(1)$ gauge fields on the NS5-branes and those of the dual
$\U(1)$ gauge fields on the D3-branes around $S^1$ in
$\CM_{p,q} \simeq S^1 \times S^3$.  Likewise, the dynamical variables
come from the holonomies of the $\SU(N)$ gauge fields on the D5-brane
(i.e.\ unshaded) regions.  The $\SU(N)$ gauge fields are dynamical and
integrated over in the path integral, except for those supported in
regions that contain part of the boundary of $\Sigma$; on the boundary
the 5-branes end on 7-branes, and boundary conditions freeze the
dynamics.  The partition function is therefore a function of the
holonomies of the nondynamical gauge fields.  Due to the $\SU(N)$
flavor symmetries, however, it only depends on these holonomies
through their conjugacy classes, which are specified by diagonal
matrices up to permutations of the entries.

As we see, this brane construction naturally leads to circle-valued
spectral parameters, though they can be analytically continued to
complex numbers.  If we wish to get complex spectral parameters
directly, instead of $\CM_{p,q}$ we can use $E_p \times_q \C$, the
fibration of $\C$ over $E_p$ such that the fiber is rotated by angle
$\Re\gamma$ or $\Im\gamma$ as it goes around either cycle of $E_p$.
The D3-branes are wrapped around $E_p$ and supported at the origin of
$\C$.  To understand the geometric meaning of the spectral parameters
in this case, we apply S-duality and then T-duality on each cycle of
$E_p$.  This chain of dualities converts the D3-branes to D1-branes
and the NS5-branes to D3-branes, all supported at points on the dual
curve $\Et_p$.  The coordinates of these points provide spectral
parameters valued in $\Et_p$.  A similar remark applies to dynamical
variables.

According to~\cite{Nieri:2015yia, Nieri:2015dts}, the partition
function on $E_p \times_q \C$ (with a Neumann boundary condition
imposed on chiral multiplets) takes the identical form as that on
$\CM_{p,q}$, except that the factor $\sqrt{pq}$ that appears in the
elliptic gamma functions are replaced with $\sqrt{q}$ and the
dynamical variables are integrated over different contours.  Thus, we
expect that this change of geometry affects neither the R-matrices
assigned to the crossings of dashed lines nor the intertwining
operators, apart from simple replacement of $\sqrt{pq}$ with
$\sqrt{q}$.

While $E_p \times_q \C$ may be a more natural choice from the point of
view of elliptic lattice models, the use of $\CM_{p,q}$ has an
advantage.  In the above construction we have placed D3-branes on the
surface $\{z_2 = 0\}$.  However, we may also wrap them on the other
distinguished surface, $\{z_1 = 0\}$.  In other words, there are
really two types of dashed lines, differing in the support of the
D3-brane.  Consequently, the Bazhanov--Sergeev model admits two sets
of L-operators related by an interchange of $p$ and $q$, leading to an
elliptic version~\cite{MR2492363} of the modular double of quantum
groups~\cite{MR1805888}.

\subsection{$\CN = 1$ supersymmetric quiver gauge theories and surface
  defects}

In practice, the supersymmetric index of the brane system discussed
above is computed in the four-dimensional effective field theory.  For
a moment, let us decompactify $\CM_{p,q}$ to~$\R^4$ and suppose there
are no D3-branes.  Since $\Sigma$ is compact, the theory that governs
the low-energy dynamics on the 5-branes is a four-dimensional theory
formulated on~$\R^4$.

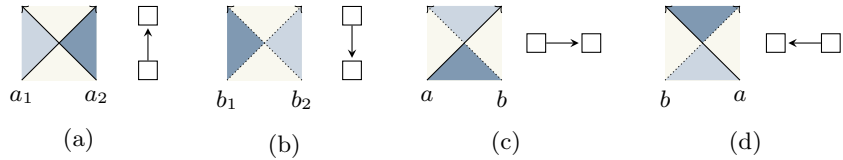
\begin{figure}
  \centering
  \subfloat[\label{fig:quiver-rule-a}]{
    \begin{tikzpicture}[scale={1/sqrt(2)}, baseline=(x.base)]
      \node (x) at (0,0) {\vphantom{x}};

      \fill[ws] (-135:1) -- (-45:1) -- (45:1) -- (135:1) -- cycle;

      \fill[dshaded] (0,0) -- (45:1) -- (-45:1) -- cycle;
      \fill[lshaded] (0,0) -- (135:1) -- (-135:1) -- cycle;

      \draw[z->] (-135:1) node[below] {$a_1$} -- (45:1);
      \draw[z->] (-45:1) node[below] {$a_2$} -- (135:1);
    \end{tikzpicture}
    \
    \begin{tikzpicture}[baseline=(x.base)]
      \node (x) at (0,0) {\vphantom{x}};

      \node[fnode, above] (z) at (0,-0.5) {};
      \node[fnode, below] (w) at (0,0.5) {};
      \draw[q->] (z) -- (w);
    \end{tikzpicture}
  }
  \quad
  \subfloat[\label{fig:quiver-rule-b}]{
    \begin{tikzpicture}[scale={1/sqrt(2)}, baseline=(x.base)]
      \node (x) at (0,0) {\vphantom{x}};

      \fill[ws] (-135:1) -- (-45:1) -- (45:1) -- (135:1) -- cycle;

      \fill[lshaded] (0,0) -- (45:1) -- (-45:1) -- cycle;
      \fill[dshaded] (0,0) -- (135:1) -- (-135:1) -- cycle;

      \draw[wz->] (-135:1) node[below] {$b_1$} -- (45:1);
      \draw[wz->] (-45:1) node[below] {$b_2$} -- (135:1);
    \end{tikzpicture}
    \
    \begin{tikzpicture}[baseline=(x.base)]
      \node (x) at (0,0) {\vphantom{x}};

      \node[fnode, above] (z) at (0,-0.5) {};
      \node[fnode, below] (w) at (0,0.5) {};
      \draw[q<-] (z) -- (w);
    \end{tikzpicture}
  }
  \quad
  \subfloat[\label{fig:quiver-rule-c}]{
    \begin{tikzpicture}[scale={1/sqrt(2)}, baseline=(x.base)]
      \node (x) at (0,0) {\vphantom{x}};

      \fill[ws] (-135:1) -- (-45:1) -- (45:1) -- (135:1) -- cycle;
    
      \fill[lshaded] (45:1) -- (0,0) -- (135:1) -- cycle;
      \fill[dshaded] (-45:1) -- (0,0) -- (-135:1) -- cycle;

      \draw[z->] (-135:1) node[below] {$a$} -- (45:1);
      \draw[wz->] (-45:1) node[below] {$b$} -- (135:1);
    \end{tikzpicture}
    \begin{tikzpicture}[baseline=(x.base)]
      \node (x) at (0,0) {\vphantom{x}};

      \node[fnode, right] (z) at (-0.5,0) {};
      \node[fnode, left] (w) at (0.5,0) {};
      \draw[q->] (z) -- (w);
    \end{tikzpicture}
  }
  \quad
  \subfloat[\label{fig:quiver-rule-d}]{
    \begin{tikzpicture}[scale={1/sqrt(2)}, baseline=(x.base)]
      \node (x) at (0,0) {\vphantom{x}};

      \fill[ws] (-135:1) -- (-45:1) -- (45:1) -- (135:1) -- cycle;

      \fill[dshaded] (45:1) -- (0,0) -- (135:1) -- cycle;
      \fill[lshaded] (-45:1) -- (0,0) -- (-135:1) -- cycle;

      \draw[wz->] (-135:1) node[below] {$b$} -- (45:1);
      \draw[z->] (-45:1) node[below] {$a$} -- (135:1);
    \end{tikzpicture}
    \begin{tikzpicture}[baseline=(x.base)]
      \node (x) at (0,0) {\vphantom{x}};

      \node[fnode, right] (z) at (-0.5,0) {};
      \node[fnode, left] (w) at (0.5,0) {};
      \draw[q<-] (z) -- (w);
    \end{tikzpicture}
  }
  \caption{The rule for assigning a quiver to a tricolor lattice.}
  \label{fig:quiver-rule}
\end{figure}

This theory is an $\CN = 1$ supersymmetric gauge theory described by a
quiver diagram drawn on $\Sigma$~\cite{Franco:2005rj, Hanany:2005ss}.
In the absence of D3-branes, the brane configuration is encoded in a
tricolor lattice on $\Sigma$.  To obtain the quiver from the lattice,
we place a node on each unshaded face and connect the nodes by arrows
according to the rule illustrated in Fig.~\ref{fig:quiver-rule}.  A
node is frozen if it is placed on a face that intersects the boundary.
Each unfrozen node represents an $\SU(N)$ gauge group, while each
frozen one represents an $\SU(N)$ flavor group.  Each arrow represents
a matter field, more precisely a chiral multiplet transforming in the
fundamental representation under the $\SU(N)$ group located at its
head and in the antifundamental representation under the $\SU(N)$
group at its tail.  The matter fields are also charged under $\U(1)$
flavor symmetries associated with NS5-branes.

Now we compactify $\R^4$ to $\CM_{p,q}$ again and compute the
partition function of the gauge theory on $\CM_{p,q}$ to obtain the
supersymmetric index.  Thanks to its protected nature, the index can
be evaluated in the weak coupling limit where the path integral
reduces to a finite-dimensional integral.  The resulting
integral~\cite{Romelsberger:2005eg} is the partition function of the
corresponding lattice model on the tricolor lattice, as defined in
section~\ref{sec:BS}.

Let us introduce a D3-brane to the system.  As we have discussed
above, it should be supported on either of the two distinguished
surfaces in $\CM_{p,q}$.  Therefore, it creates a surface defect in
the four-dimensional theory.  In flat spacetime, this defect would
preserve half of the supercharges that generate $\CN = (0,2)$
supersymmetry on the surface.

The introduction of the surface defect changes the supersymmetric
index.  One way to determine its effect is to realize the defect as a
two-dimensional theory coupled to the four-dimensional theory, and
compute the index of the coupled system in the weak coupling limit.
It turns out that the defect is represented by a difference operator
acting on $\SU(N)$ gauge or flavor fugacities~\cite{Gadde:2013ftv}.
In the lattice model, the action of this operator is induced by the
dashed line inserted by the D3-brane.

\begin{figure}
  \centering
  \subfloat[\label{fig:class-Sk-quiver}]{
    \begin{tikzpicture}[baseline=(x), scale=0.8]
      \node[fnode] (a0) at (0,0) {};
      \node[fnode] (b0) at (0,-1) {};
      \node[fnode] (c0) at (0,-2) {};
      \node[gnode] (a1) at (1,0) {};
      \node[gnode] (b1) at (1,-1) {};
      \node[gnode] (c1) at (1,-2) {};
      \node[gnode] (a2) at (2,0) {};
      \node[gnode] (b2) at (2,-1) {};
      \node[gnode] (c2) at (2,-2) {};
      
      \begin{scope}[shift={(0.7,0)}]
        \node[gnode] (a3) at (3,0) {};
        \node[gnode] (b3) at (3,-1) {};
        \node[gnode] (c3) at (3,-2) {};
        \node[fnode] (a4) at (4,0) {};
        \node[fnode] (b4) at (4,-1) {};
        \node[fnode] (c4) at (4,-2) {};
      \end{scope}
      
      \draw[q->] (a1) -- (b0);
      \draw[q->] (b1) -- (c0);
      
      \draw[q->] (a2) -- (b1);
      \draw[q->] (b2) -- (c1);
      
      \draw[q->] (a4) -- (b3);
      \draw[q->] (b4) -- (c3);
      
      \draw[q->] (a0) -- (a1);
      \draw[q->] (a1) -- (a2);
      \draw[q->] (a3) -- (a4);
      
      \draw[q->] (b0) -- (b1);
      \draw[q->] (b1) -- (b2);
      \draw[q->] (b3) -- (b4);
      
      \draw[q->] (c0) -- (c1);
      \draw[q->] (c1) -- (c2);
      \draw[q->] (c3) -- (c4);
      
      \draw[q->] (b1) -- (a1);
      \draw[q->] (c1) -- (b1);
      
      \draw[q->] (b2) -- (a2);
      \draw[q->] (c2) -- (b2);
      
      \draw[q->] (b3) -- (a3);
      \draw[q->] (c3) -- (b3);
      
      \draw[q->] ($(a0) + (0.5,0.5)$) -- (a0);
      \draw[q->] ($(a1) + (0.5,0.5)$) -- (a1);
      \draw[q->] ($(a2) + (0.5,0.5)$) -- (a2);
      \draw[q->] ($(a3) + (0.5,0.5)$) -- (a3);
      \draw[q->] ($(b2) + (0.5,0.5)$) -- (b2);
      \draw[q->] ($(c2) + (0.5,0.5)$) -- (c2);
      
      \draw[q-] (a3) -- ($(a3) + (-0.5,-0.5)$);
      \draw[q-] (b3) -- ($(b3) + (-0.5,-0.5)$);
      \draw[q-] (c1)  -- ($(c1) + (-0.5,-0.5)$);
      \draw[q-] (c2)  -- ($(c2) + (-0.5,-0.5)$);
      \draw[q-] (c3)  -- ($(c3) + (-0.5,-0.5)$);
      \draw[q-] (c4)  -- ($(c4) + (-0.5,-0.5)$);
      
      \draw[q-] (a1) -- ($(a1) + (0,0.5)$);
      \draw[q-] (a2) -- ($(a2) + (0,0.5)$);
      \draw[q-] (a3) -- ($(a3) + (0,0.5)$);
      
      \draw[q->] ($(c1) + (0,-0.5)$) -- (c1);
      \draw[q->] ($(c2) + (0,-0.5)$) -- (c2);
      \draw[q->] ($(c3) + (0,-0.5)$) -- (c3);
      
      \draw[q-] (a2) -- ($(a2) + (0.5,0)$);
      \draw[q-] (b2) -- ($(b2) + (0.5,0)$);
      \draw[q-] (c2) -- ($(c2) + (0.5,0)$);
      
      \draw[q->] ($(a3) + (-0.5,0)$) -- (a3);
      \draw[q->] ($(b3) + (-0.5,0)$) -- (b3);
      \draw[q->] ($(c3) + (-0.5,0)$) -- (c3);
      
      \node (x) at (2.9,-1) {$\dots$};
    \end{tikzpicture}
  }
  \qquad
  \subfloat[\label{fig:class-Sk-sd}]{
    \begin{tikzpicture}[baseline=(x.base), scale=0.8]
      \node (x) at (0,1.5) {\vphantom{x}};
      
      \fill[ws] (0,0) rectangle (2.5,2.7);
      
      \begin{scope}
        \fill[lshaded] (1,0.7) rectangle (2.5,1);
        \fill[dshaded] (0,0) -- (1,0) -- (1,0.7) -- (0.7,0.7)
        -- (0.7,0.3) -- (0,0.3) -- cycle;
        
        \draw[z->] (0,0.3) node[left] {$a_{f+1}$}
        -- (0.7,0.3) -- (0.7,0.7) -- (2.5,0.7);
        
        \node (zi+1) at (2.05,0.35) {$z_{f+1}$};
      \end{scope}
      
      \begin{scope}[shift={(0,1)}]
        \fill[lshaded] (1,0.7) rectangle (2.5,1);
        \fill[dshaded] (0,0) -- (1,0) -- (1,0.7) -- (0.7,0.7)
        -- (0.7,0.3) -- (0,0.3) -- cycle;
        
        \draw[z->] (0,0.3) node[left] {$a_f$}
        -- (0.7,0.3) -- (0.7,0.7) -- (2.5,0.7);
        
        \node (zi+1) at (2.05,0.35) {$z_f$};
      \end{scope}
      
      \begin{scope}[shift={(0,2)}]
        \fill[dshaded] (0,0) -- (1,0) -- (1,0.7) -- (0.7,0.7)
        -- (0.7,0.3) -- (0,0.3) -- cycle;
        
        \draw[z->] (0,0.3) node[left] {$a_{f-1}$}
        -- (0.7,0.3) -- (0.7,0.7);
        
        \node (zi+1) at (2.05,0.35) {$z_{f-1}$};
      \end{scope}
      
      \draw[wz->] (1,0) -- (1,2.7);
      \draw[wz->] (0,1) node[left] {$b_{f+1}$} -- (2.5,1);
      \draw[wz->] (0,2) node[left] {$b_f$} -- (2.5,2);
      \draw[dr->] (1.5,2.7) node[above] {$c$} -- (1.5,0);

      \draw[boundary] (2.5,0) -- (2.5,2.7);
    \end{tikzpicture}
  }
  \caption{(a) A quiver describing a class-$\CS_k$ theory, placed on a
    finite-length cylinder.  (b) A surface defect inserted near the
    right end of the cylinder.}
  \label{fig:class-Sk}
\end{figure}
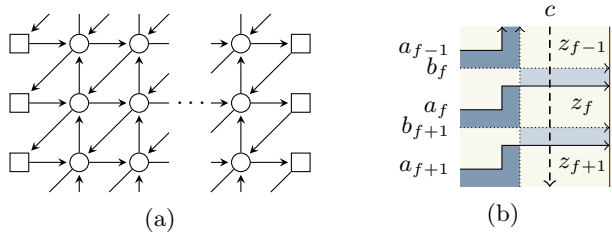

To be specific, let us take $\Sigma$ to be a finite-length cylinder
and consider a lattice that leads to the quiver shown in
Fig.~\ref{fig:class-Sk-quiver}.  The vertical direction is periodic,
and there are $k$ nodes in each column.  This theory is a $\Z_k$
orbifold of an $\CN = 2$ supersymmetric gauge theory described by a
linear quiver, and is a basic example of a class-$\CS_k$
theory~\cite{Gaiotto:2015usa}.  The supersymmetric index of the
theory, or equivalently the partition function of the lattice model,
is a function of the dynamical variables associated with the $k$
$\SU(N)$ flavor groups on the right end of the quiver, which we denote
by $z_f = (z_{f,1}, \dotsc, z_{f,N})$, $f=1$, $\dotsc$, $k$, as well
as those associated with the $k$ $\SU(N)$ flavor groups on the left
end.

In this theory, we introduce a surface defect that inserts a dashed
line in the lattice as in Fig.~\ref{fig:class-Sk-sd}.  Let us define
an L-operator $\LFt$ by the crossing of a dashed line and a double
line in an unshaded background.  More explicitly, it has matrix
elements
\begin{equation}
  \label{eq:LFt}
  \LFt(a,b)^j_i
  =
  \Delta_{w,j}^{-1}
  \LFt(a,b; z, w)^j_i
  \Delta_{z,i}^{-1}
  \,,
\end{equation}
with $\LFt(a,b; z, w)$ given by
\begin{equation}
  \label{eq:LFt-g}
  \LFt\Bigl(\frac{c}{a},\frac{c}{b}; z, w\Bigr)
  =
  \begin{tikzpicture}[scale=0.6, baseline=(x.base)]
    \node (x) at (1,1) {\vphantom{x}};

    \fill[ws] (0,0) rectangle (2,2);
    \draw[dtr->] (1,0) node[below] {$(a,b)$} -- (1,2);
    \draw[dr->] (0,1) node[left] {$c$} -- (2,1);

    \node at (0.5,1.5) {$z$};
    \node at (1.5,0.5) {$w$};
  \end{tikzpicture}
  \ =
  \begin{tikzpicture}[scale=0.6, baseline=(x.base)]
    \node (x) at (0,1) {\vphantom{x}};

    \fill[ws] (0,0) rectangle (0.75,2);
    \fill[ws] (1.25,0) rectangle (2,2);
    \fill[lshaded] (0.75,2) rectangle (1.25,0);

    \draw[dr->] (0,1) node[left] {$c$} -- (2,1);
    \draw[wz->] (0.75,0) node[below] {$b$}-- (0.75,2);
    \draw[z->] (1.25,0) node[below] {$a$} -- (1.25,2);

    \node at (0.325,1.5) {$z$};
    \node at (1.625,0.5) {$w$};
  \end{tikzpicture}
  \ =
  \St'\Bigl(\frac{c}{a}; w\Bigr) \St\Bigl(\frac{c}{b}; z\Bigr)
  \,.
\end{equation}
Just as $\LF$ does, this L-operator gives a representation of
$E_{\tau, \gamma/2}(\slf_N)$.  Looking at Fig.~\ref{fig:class-Sk-sd},
we conclude that the surface defect acts on the index as the transfer
matrix
\begin{equation}
  \Tr\Bigl(\LFt_k\Bigl(\frac{c}{a_k}, \frac{c}{b_k}\Bigr)
  \dotsb
  \LFt_1\Bigl(\frac{c}{a_1}, \frac{c}{b_1}\Bigr)\Bigr)
  \,,
\end{equation}
where $\LFt_f$ is the copy of $\LFt$ corresponding to the crossing of
the dashed line and the $f$th pair of dotted and solid lines.

A simple calculation yields
\begin{equation}
  \LFt\Bigl(\frac{c}{a},\frac{c}{b}\Bigr)^j_i
  =
  \frac{\theta\bigl(d^N \sqrt{p/q} aw_j/bz_i\bigr)}
       {\theta(d^N)}
  \dprod_{k (\neq i)}
  \frac{\theta(\sqrt{p/q} aw_j/bz_k)}
       {\theta(z_k/z_i)}
  \Delta_{z,i}^{-1} \Delta_{w,j}^{-1}
  \,,
\end{equation}
with $d^{-1} = q^{-1/N} \sqrt{pq} c/b$.  For $N = 2$, the transfer
matrix resulting from this formula reproduces the difference operator
obtained in~\cite{Maruyoshi:2016caf}, up to an overall factor that is
independent of the dynamical variables.  For $N > 2$, the transfer
matrix is consistent with the result found in~\cite{Ito:2016fpl},
where the action of surface defects in the symmetric powers of the
vector representation was computed for restricted values of the
spectral parameters.

\subsection{Fusion procedure and exterior powers of the vector
  representation}

So far we have considered a particular type of surface defect that is
created by a single D3-brane ending on the 5-branes.  In fact, this is
the simplest member of a family of surface defects that can be
constructed by similar brane configurations, which in general contain
multiple D3-branes and extra NS5-branes.  The possible types of these
defects are in one-to-one correspondence with the irreducible
finite-dimensional representations of $\SU(N)$.  What we have
discussed above is the case of the vector representation.

Correspondingly, the lattice model should admit a family of dashed
lines labeled by the irreducible finite-dimensional representations of
$\SU(N)$, not just the one that we have used so far.  Indeed, there is
a method to generate such a family.  This is the fusion procedure,
initiated by Kulish, Reshetikhin and Sklyanin~\cite{Kulish:1981gi} in
the context of the rational analog of the Belavin model.
In~\cite{Kulish:1981gi}, the R-matrices for the symmetric and exterior
powers of the vector representation were obtained by this procedure.
Subsequently, the construction was generalized by
Cherednik~\cite{MR839706} to arbitrary irreducible representations at
the full elliptic level, and applied to the Jimbo--Miwa--Okado model
in~\cite{Jimbo:1988gs}.%
\footnote{In connection with surface defects, it should also be noted
  that the transfer matrices in the symmetric powers of the vector
  representation are residues of a certain integral operator
  \cite{Gaiotto:2012xa, Gaiotto:2015usa}.  In
  \cite{Chicherin:2014dya}, the general fused L-operator were obtained
  for the elliptic modular double for $N = 2$ via reduction of the
  BSDS R-operator.}

Here we explain the fusion procedure for the exterior powers of the
vector representation $V = \C^N$, following the treatment
in~\cite{MR1606760}.  Concretely, we construct an L-operator
$\LFt_{\wedge^n V}$ that is associated with the $n$th exterior power
$\wedge^n V$ and reduces for $n = 1$ to the L-operator $\LFt$ defined
in~\eqref{eq:LFt} and~\eqref{eq:LFt-g}.

Let $J_n(V)$ be the subspace of $V^{\otimes n}$ consisting of tensors
of the form $\sigma v - \sgn(\sigma) v$, where $\sigma \in S_n$ is a
permutation of $n$ elements.  Then $\wedge^n V$ is the quotient
$V^{\otimes n}/J_n(V)$.  We set
\begin{equation}
  \RFb(u, \lambda)
  =
  -\frac{\theta_1(u+\gamma)}{2\theta_1(\gamma)}
  \RF(u, \lambda)
  \,.
\end{equation}
From the definition of $\RF$, it is easy to check
$J_2(V) \subset \ker \RFb(-\gamma,\lambda)$.  We actually have
$\ker \RFb(-\gamma,\lambda) = J_2(V)$ for generic values of $\lambda$,
as can be seen from the fact that $\RFb(-\gamma,\lambda)$ acts on
antisymmetric tensors as the identity in the limit $\gamma \to 0$.

To implement the fusion procedure, we introduce an operator defined by
the following diagram:
\begin{equation}
  \Biggl(\prod_{1 \leq i < j \leq N}
  -\frac{\theta_1(u_i - u_j +\gamma)}{2\theta_1(\gamma)}
  \Biggr)
  \times
  \begin{tikzpicture}[scale=0.6, baseline=(x.base)]
    \node (x) at (1,0.8) {\vphantom{x}};

    \fill[ws] (0,-0.5) rectangle (3,2.25);

    \node at (0.25,0.8) {$\vdots$};
    \node at (2.75,1.3) {$\vdots$};

    \draw[dr->] (0,1.75) node[left] {$u_1$} -- (0.5,1.75) -- (2,0) --
    (3,0) node[right] {$u_1$};

    \draw[dr->] (0,1.25) node[left] {$u_2$} -- (1.5,1.25) --
    ({1.5+0.75*1.5/1.75},0.5) -- (3,0.5) node[right] {$u_2$};

    \draw[dr->] (0,0) node[left] {$u_n$} -- (1,0) -- (2.5,1.75) --
    (3,1.75) node[right] {$u_n$};
  \end{tikzpicture}
  .
\end{equation}
In the middle of the diagram we have a composition of $n(n-1)/2$
crossings that generate the permutation
$(1, 2,\dotsc, n) \mapsto (n,\dotsc, 2, 1)$; the details are not
important since the Yang--Baxter equation guarantees that any two such
compositions produce the same operator.  The prefactor converts the
R-matrices $\RF_{ij}$ that appear in the diagram to $\RFb_{ij}$.
Since any permutation is a product of adjacent transpositions,
$J_n(V)$ is spanned by tensors of the form $v + P_{i, i+1} v$, with
$P_{i,i+1}$ being the swap of the $i$th and $(i+1)$th factors.
Furthermore, for any adjacent pair $(i,i+1)$, we can find a
representation of the above operator such that $\RFb_{i,i+1}$ appears
in the leftmost position in the diagram.  It follows that the kernel
of this operator, evaluated at
$(u_1,\dotsc, u_n) = (u, u+\gamma, \dotsc, u + (n-1)\gamma)$, is
generically $J_n(V)$.

Now we consider the following tensor product of $n$ copies of $\LFt$:
\begin{equation}
  \begin{tikzpicture}[scale=0.6, baseline=(x.base)]
    \node (x) at (0,0.8) {\vphantom{x}};

    \fill[ws] (0,-0.5) rectangle (2,2.25);
    \fill[lshaded] (0.75,-0.5) rectangle (1.25,2.25);

    \draw[wz->] (0.75,-0.5) node[below] {$a$}-- (0.75,2.25);
    \draw[z->] (1.25,-0.5) node[below] {$b$} -- (1.25,2.25);

    \node at (0.25,0.8) {$\vdots$};
    \node at (1.75,0.8) {$\vdots$};

    \draw[dr->] (0,1.75) node[left] {$c$} -- (2,1.75);
    \draw[dr->] (0,1.25) node[left] {$q^{1/N} c$} -- (2,1.25);
    \draw[dr->] (0,0) node[left] {$q^{(n-1)/N} c$} -- (2,0);
  \end{tikzpicture}
  \ .
\end{equation}
The relation
\begin{equation}
  \label{eq:LFtn}
  \begin{tikzpicture}[scale=0.6, baseline=(x.base)]
    \node (x) at (0,0.8) {\vphantom{x}};

    \fill[ws] (0,-0.5) rectangle (4.5,2.25);
    \fill[lshaded] (0.75,-0.5) rectangle (1.25,2.25);

    \draw[wz->] (0.75,-0.5) -- (0.75,2.25);
    \draw[z->] (1.25,-0.5) -- (1.25,2.25);

    \node at (0.25,0.8) {$\vdots$};
    \node at (1.75,0.8) {$\vdots$};
    \node at (4.25,1.3) {$\vdots$};

    \draw[dr->] (0,1.75) -- (2,1.75) -- (3.5,0) --  (4.5,0);

    \draw[dr->] (0,1.25) -- (3,1.25) -- ({3+0.75*1.5/1.75},0.5) --
    (4.5,0.5);

    \draw[dr->] (0,0) -- (2.5,0) -- (4,1.75) -- (4.5,1.75);
  \end{tikzpicture}
  \ = \
  \begin{tikzpicture}[scale=0.6, baseline=(x.base)]
    \node (x) at (0,0.8) {\vphantom{x}};

    \fill[ws] (0,-0.5) rectangle (4.5,2.25);
    \fill[lshaded] (3.25,-0.5) rectangle (3.75,2.25);

    \draw[wz->] (3.25,-0.5) -- (3.25,2.25);
    \draw[z->] (3.75,-0.5) -- (3.75,2.25);

    \node at (0.25,0.8) {$\vdots$};
    \node at (2.75,1.3) {$\vdots$};
    \node at (4.25,1.3) {$\vdots$};
    
    \draw[dr->] (0,1.75) -- (0.5,1.75) -- (2,0) -- (4.5,0);

    \draw[dr->] (0,1.25) -- (1.5,1.25) -- ({1.5+0.75*1.5/1.75},0.5) --
    (4.5,0.5);

    \draw[dr->] (0,0) -- (1,0) -- (2.5,1.75) -- (4.5,1.75);
  \end{tikzpicture}
\end{equation}
shows that this operator leaves $J_n(V)$ invariant.  Hence, it
descends to a well-defined operator $\LFt_{\wedge^n V}(c/a, c/b)$
under the projection $V^{\otimes n} \to \wedge^n V$.  This is the
desired L-operator.

It is clear that the same argument holds when the pair of vertical
lines in~\eqref{eq:LFtn} is replaced with any line that satisfies the
Yang--Baxter equation with two dashed lines.  Therefore, it makes
sense to define a dashed line in the representation $\wedge^n V$ as
the image of the $n$ copies of dashed lines in the picture under the
projection:
\begin{equation}
  \begin{tikzpicture}[scale=0.6, baseline=(x.base)]
    \node (x) at (0.5,0) {\vphantom{x}};
    \draw[dr->] (0,0) node[left] {$\wedge^n V$, $c$} -- (2,0);
  \end{tikzpicture}
  \
  =
  \left[
  \begin{tikzpicture}[scale=0.6, baseline=(x.base)]
    \node (x) at (0,0.8) {\vphantom{x}};

    \node at (1,0.8) {$\vdots$};

    \draw[dr->] (0,1.75) node[left] {$c$} -- (2,1.75);
    \draw[dr->] (0,1.25) node[left] {$q^{1/N} c$} -- (2,1.25);
    \draw[dr->] (0,0) node[left] {$q^{(n-1)/N} c$} -- (2,0);
  \end{tikzpicture}
  \right]
  \,.
\end{equation}

The fusion procedure just described should reflect the structure of
the brane construction of a surface defect in the representation
$\wedge^n V$.  In this construction, we start with $n$ D3-branes
ending on a stack of $N$ coincident D5-branes.  These D3-branes create
$n$ dashed lines, each in the vector representation.  To this
configuration we add an NS5-brane, and let the other ends of the
D3-branes terminate on it; see Fig.~\ref{fig:D3-ext}.  The support of
the NS5-brane is chosen in such a way that the D3-branes necessarily
become coincident.  The exclusion principle for branes then forces the
D3-branes to end on separate D5-branes in the stack.  There are
$N!/n!(N-n)!$ different ways in which the D3-branes end on the
D5-branes, and they correspond to the standard basis vectors of
$\wedge^n V$.

\begin{figure}
  \centering
  \begin{tikzpicture}
      \draw (0,-1) -- (0,1) node[below left] {$N$ D5};
      \draw (0.1,-1) -- (0.1,1);
      \draw (0.2,-1) -- (0.2,1);
      \draw (0.3,-1) -- (0.3,1);

      \draw (0,0.1) -- (2,0.1);
      \draw (0.1,0) -- (2,0);
      \draw (0.2,-0.1) -- (2,-0.1);

      \node[anchor=south, above] at (1,0.1) {$n$ D3};

      \draw[fill=white] (2,0) circle [radius=0.2];
      \draw ($(2,0) + (45:0.2)$) -- ++(-135:0.4);
      \draw ($(2,0) + (-45:0.2)$) -- ++(135:0.4);

      \node[anchor=north west, below] at (2,-0.25) {NS5};
    \end{tikzpicture}
    \caption{The brane configuration for a surface defect in the
      representation $\wedge^n V$.  Here, $(N,n) = (4,3)$ and the
      dashed line carries the state $e_1 \wedge e_2 \wedge e_3$}
  \label{fig:D3-ext}
\end{figure}

In order to check the correspondence between the brane construction
and the fusion procedure more quantitatively, let us consider the
situation in Fig.~\ref{fig:class-Sk-sd} again.  If we replace the
dashed line in the picture with one in the representation
$\wedge^n V$, then the new line acts on the lattice model as a
transfer matrix consisting of $k$ copies of $\LFt_{\wedge^n V}$.
According to our proposal, it should coincide with the difference
operator that represents the action of the corresponding surface
defect on the index of the $\CN = 1$ supersymmetric gauge theory
described by the quiver in Fig.~\ref{fig:class-Sk-quiver}.  The gauge
theory result is known for $k = 1$~\cite{Bullimore:2014nla}, so let us
calculate the transfer matrix in this case.

Let $I$ run over the subsets of $\{1, \dotsc, N\}$ of order $n$, and
$e_I = e_{i_1} \wedge \dotsb \wedge e_{i_n}$ be a basis vector of
$\wedge^n V$, where $i_r$ stands for the $r$th smallest number in $I$.
We define
\begin{equation}
  \label{eq:LFt-wedge}
  \LFt_{\wedge^n V}\Bigl(\frac{c}{a},\frac{c}{b}; z, w\Bigr)
  =
  \begin{tikzpicture}[scale=0.6, baseline=(x.base)]
    \node (x) at (1,1) {\vphantom{x}};

    \fill[ws] (0,0) rectangle (2,2);
    \draw[dtr->] (1,0) node[below] {$(a,b)$} -- (1,2);
    \draw[dr->] (0,1) node[left] {$\wedge^n V$, $c$} -- (2,1);

    \node at (0.5,1.5) {$z$};
    \node at (1.5,0.5) {$w$};
  \end{tikzpicture}
  \ .
\end{equation}
Then, the transfer matrix we wish to calculate is
\begin{equation}
  \label{eq:TrLFwedge}
  \Tr\LFt_{\wedge^n V}\Bigl(\frac{c}{a},\frac{c}{b}\Bigr)
   =
  \sum_I
  \LFt_{\wedge^n V}\Bigl(\frac{c}{a},\frac{c}{b}; z, T_I^{-1} z\Bigr)^I_I
  T_I^{-1}
  \,,
\end{equation}
with $T_I = \prod_{i \in I} T_i$.  The matrix-valued
function~\eqref{eq:LFt-wedge} satisfies an RLL relation with a
generalization of Felder's R-matrix.

In terms of intertwining operators, the coefficient function in front
of $T_I^{-1}$ is given by
\begin{multline}
  \label{eq:LftII}
  \LFt_{\wedge^n V}\Bigl(\frac{c}{a},\frac{c}{b}; z, T_I^{-1} z\Bigr)^I_I
  \\
  =
  \sum_{\sigma \in S_n} \sgn(\sigma)
  \prod_{r=1}^n
  \biggl[
  \St'\biggl(q^{(r-1)/N} \frac{c}{a}, T_{\sigma \cdot I_r}^{-1} z \biggr)
  \St\biggl(q^{(r-1)/N} \frac{c}{b}, T_{I_{r-1}}^{-1} z \biggr)
  \biggr]^{i_{\sigma(r)}}_{i_r}
  \,,
\end{multline}
where $I_{r-1} = \{i_1, \dotsc, i_{r-1}\}$ and
$\sigma \cdot I_r = \{i_{\sigma(1)}, \dotsc, i_{\sigma(r)}\}$.  Using
identity~\eqref{eq:St'=St'} and the formula~\cite{MR1463830}
\begin{multline}
  \sum_{\sigma \in S_n} \sgn(\sigma)
  \prod_{r=1}^n
  \Biggl[
  \Phi\biggl(v + (r-1)\gamma,
  \lambda - \gamma \sum_{s=1}^{r-1} \omega_{i_s}\biggr)^{-1}
  \Phi\biggl(u + (r-1)\gamma,
  \lambda - \gamma \sum_{s=1}^{r-1} \omega_{i_{\sigma(s)}}\biggr)
  \Biggr]^{i_r}_{i_{\sigma(r)}}
  \\
  =
  \frac{\theta_1\bigl(v + (u-v)n/N\bigr)}{\theta_1(v)}
  \prod_{\substack{i \in I \\ k \notin I}}
  \frac{\theta_1\bigl(\lambda_{ki} + (u-v)/N\bigr)}{\theta_1(\lambda_{ki})}
  \,,
\end{multline}
we can rewrite it as
\begin{equation}
  \LFt_{\wedge^n V}\Bigl(\frac{c}{a},\frac{c}{b}; z, T_I^{-1} z\Bigr)^I_I
  =
  \Bigl(\sqrt{pq} \frac{a}{b}\Bigr)^{n(n-1)/2}
  \frac{\theta\bigl(d^N (\sqrt{p/q} a/b)^n\bigr)}
       {\theta(d^N)}
  \prod_{\substack{i \in I \\ k \notin I}}
  \frac{\theta(\sqrt{p/q} az_i/bz_k)}
       {\theta(z_k/z_i)}
  \,.
\end{equation}
Here $d^{-1} = q^{-1/N} \sqrt{pq} c/b$ as before.

Up to a factor independent of the dynamical variable, the transfer
matrix computed above reproduces the difference operator that was
proposed in~\cite{Bullimore:2014nla} to represent the action of the
surface defect on the supersymmetric index.  (See section~4.1
of~\cite{Maruyoshi:2016caf} for the relation between the spectral
parameters used here and the fugacities used
in~\cite{Bullimore:2014nla}.)  Varying $n$ from $1$ to $N$, we obtain
a family of mutually commuting difference operators from the transfer
matrix.  As was shown in~\cite{MR1190749}, these are related by
conjugation by a function to the Ruijsenaars
operators~\cite{Ruijsenaars:1986pp}, the Hamiltonians of the elliptic
Ruijsenaars--Schneider model of type $A_{N-1}$.

\section*{Acknowledgments}

I would like to thank Kevin Costello for illuminating discussions and
directing my attention to Felder's dynamical R-matrix, and Vladimir
Bazhanov and Sergey Sergeev for explaining to me the relevance
of~\cite{Sergeev:1992ap} to their model.  I am also grateful to Piotr
Su\l kowski and Petr Va\v sko for their comments.  This work is
supported by the ERC Starting Grant No.~335739 ``Quantum fields and
knot homologies'' funded by the European Research Council under the
European Union's Seventh Framework Programme.

\end{document}